\documentclass[
aps,prd,twocolumn,reprint,amsmath,amssymb,preprintnumbers,superscriptaddress,nobibnotes,nofootinbib,floatfix,longbib
]{revtex4-1}

\usepackage{graphicx}
\usepackage{dcolumn}
\usepackage{bm}
\usepackage{color}
\usepackage{amssymb}
\usepackage{amsmath}
\usepackage{braket}
\usepackage[colorlinks=true,allcolors=Blue,pdfusetitle]{hyperref}
\usepackage{rotating}
\usepackage{multirow}
\usepackage{graphicx}
\usepackage[dvipsnames]{xcolor}
\usepackage{svg}
\usepackage{esdiff}
\usepackage[normalem]{ulem}

\definecolor{Green}{RGB}{0, 128, 0}
\newcommand{\orcid}[1]{\href{https://orcid.org/#1}{\includegraphics[width=10pt]{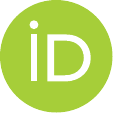}}}

\usepackage{tikz}
\usetikzlibrary{quantikz}

\begin{document}
\preprint{LLNL-JRNL-2004731-DRAFT}
\title{Advancing quantum simulations of nuclear shell model with noise-resilient protocols}

\author{Nifeeya Singh \orcid{0000-0002-0845-4221}}
\email{n\_singh@ph.iitr.ac.in}
\affiliation{%
 Department of Physics, Indian Institute of Technology Roorkee, Roorkee
247 667, India 
}%

\author{Pooja Siwach \orcid{0000-0001-6186-0555}}
\email{siwach1@llnl.gov}
\affiliation{Nuclear and Chemical Science Division, Lawrence Livermore National Laboratory, Livermore, California 94551, USA}

\author{P. Arumugam \orcid{0000-0001-9624-8024}}%
 \email{arumugam@ph.iitr.ac.in}
\affiliation{%
 Department of Physics, Indian Institute of Technology Roorkee, Roorkee
247 667, India 
}%

\date{\today}

\begin{abstract}
 \begin{description}
\item[Background] Some of the computational limitations in solving the nuclear many-body problem could be overcome by utilizing quantum computers. The nuclear shell-model calculations providing deeper insights into the properties of atomic nuclei, is one such case with high demand for resources as the size of the Hilbert space grows exponentially with the number of particles involved. Quantum algorithms are being developed to overcome these challenges and advance such calculations.
\item[Purpose] To develop quantum circuits for the nuclear shell-model, leveraging the capabilities of noisy intermediate-scale quantum (NISQ) devices. We aim to minimize resource requirements (specifically in terms of qubits and gates) and strive to reduce the impact of noise by employing relevant mitigation techniques.
\item[Methods] We achieve noise resilience by designing an optimized ansatz for the variational quantum eigensolver (VQE) based on Givens rotations and incorporating qubit-ADAPT-VQE in combination with variational quantum deflation (VQD) to compute ground and excited states incorporating the zero-noise extrapolation mitigation technique. Furthermore, the qubit requirements are significantly reduced by mapping the basis states to qubits using Gray code encoding and generalizing transformations of fermionic operators to efficiently represent many-body states.  
\item[Results] By employing the noise-resilient protocols, we achieve the ground and excited state energy levels of $^{38}$Ar and $^6$Li  with better accuracy. These energy levels are presented for noiseless simulations, noisy conditions, and after applying noise mitigation techniques. Results are compared for Jordan Wigner and Gray code encoding using VQE, qubit-ADAPT-VQE, and VQD.
\item[Conclusions] Our work highlights the potential of noise-resilient protocols to leverage the full potential of NISQ devices in scaling the nuclear shell model calculations, offering a pathway toward more complex quantum simulations in nuclear physics. This approach establishes a framework for studying other nuclear systems with improved quantum resource efficiency, marking a significant advancement in applying quantum computing to realistic nuclear physics applications.
\end{description}

\end{abstract}

\maketitle

\section{INTRODUCTION}\label{intro}
Quantum computing has the potential to overcome some of the limitations of classical (conventional) computing involving many-body systems like atomic nuclei. For example, the application of nuclear shell-model that can provide deeper insights into the structure of nuclei is often limited to very light nuclei and with over-simplifying approximations to describe heavier nuclei. These limitations ascribed to the exponential increase in the Hilbert space with an increase in the number of nucleons could be handled better with quantum computing owing to its inherent parallelism and efficient algorithms estimating the expectation values. Several studies have been carried out in nuclear physics and quantum many-body problems in general~\cite{Dumitrescu:2018,Roggero:2019,Roggero:2020,Roggero1:2020,Siwach:PRC2021,Siwach:PRA2021,Denis:2020,Denis:2022,Obiol:sr2023,Chandan:prc2023,Bhoy:njp2024}, with the help of quantum computers.

Due to the limited capabilities of current noisy intermediate-scale quantum (NISQ) devices, the hybrid quantum-classical algorithms are the most suitable to perform quantum simulations of many-body systems. The variational quantum algorithms including the variational quantum eigensolvers (VQE)~\cite{VQE_nature1,Siwach:PRC2021} are among the most promising algorithms for simulating the complex many-body problems. The standard VQE approach involves several stages, such as mapping physical degrees of freedom (e.g., fermionic operators) to qubits, preparing an initial reference state, conducting variational optimization, evaluating operator expectation values (particularly the Hamiltonian), and applying techniques for error mitigation. While VQE has been successfully applied to various systems, it comes with inherent limitations. One major challenge is the circuit depth.
In many cases, the ansatz used in VQE, such as the unitary coupled-cluster singles and doubles (UCCSD) ansatz~\cite{ucc1,ucc2,ucc3}, requires deep circuits with a large number of variational parameters. This can be computationally expensive and impractical for NISQ devices, which are constrained by smaller coherence times and larger gate fidelities. Additionally, the process of choosing an appropriate ansatz remains a key challenge, as the form of the ansatz significantly impacts the accuracy and complexity of the calculations. If the ansatz does not adequately represent the ground-state subspace or the relevant symmetries, the VQE method may fail to converge to the correct solution.

To address these issues, first, we explore the application of VQE in the shell model calculations by designing an optimized version of the particle-number conserving ansatz based on Given rotation~\cite{Arrazola:quantum2022}, significantly reducing the single- and two-qubit gates. We further explore the qubit-ADAPT-VQE algorithm~\cite{Tang:PRXQuantum2021}. Unlike traditional VQE, which relies on a fixed ansatz, qubit-ADAPT-VQE constructs the ansatz dynamically, iteratively adding operators that most significantly influence the energy. This process reduces the number of variational parameters required and avoids the complexity of including unnecessary operators, which is often the case with traditional methods like UCCSD. Qubit-ADAPT-VQE operates with a qubit-based operator pool, making it possible to avoid the costly fermion-to-spin mapping that typically leads to high gate overhead in fermionic-ADAPT~\cite{Grimsley:nc2019,adapt}. By selecting operators that contribute most to the energy at each iteration, this dynamic approach ensures that the ansatz is tailored specifically to the system, optimizing both computational efficiency and accuracy. In the present work, we further integrate qubit-ADAPT-VQE with variational quantum deflation (VQD)~\cite{Higgott:quantum2019} to calculate the excited state energies.

Even with a powerful technique like qubit-ADAPT-VQE, the resource requirement is still large for lighter nuclei (e.g., $^6$Li and $^{38}$Ar considered in the present work). Therefore, we implement the Gray code encoding to map our system on qubits~\cite{Siwach:PRC2021,dimatteo2021} by significantly reducing the number of qubits required. In the previous studies~\cite{Siwach:PRC2021,dimatteo2021}, the transformation of fermionic operators is carried out without considering the antisymmetrization of wave function since the system under consideration had only one valence particle. We generalize these transformations for a many-body case by properly incorporating the antisymmetrization of wave functions.

The paper is organized as follows. Section~\ref{sec:theoretical} introduces the nuclear Hamiltonian, including its encoding and the construction of particle conserving ansatz. Section~\ref{sec:sec2} presents the details about the applied quantum algorithms. Section~\ref{sec:Results} discusses the results for $^{38}$Ar and $^{6}$Li under both encoding schemes, followed by zero noise extrapolation. Finally, the conclusions derived from this work are presented in Section~\ref{sec:Conclusions}.  

\section{Nuclear Hamiltonian}\label{sec:theoretical}
In this section, the nuclear Hamiltonian $H$ is constructed within the shell model~\cite{sm1,sm2,sm3} framework to calculate the energy levels of a chosen nucleus. The shell model Hamiltonian $H$ comprises two components -- the single-particle energy term and the two-body interaction term, expressed as~\cite{suhonen2007nucleons}

\begin{equation}\label{eq:H}
    H = \sum_{\alpha} \epsilon_{\alpha} c_{\alpha}^\dagger c_{\alpha} + \frac{1}{4} \sum_{\alpha \beta \gamma \delta} \overline{v}_{\alpha \beta \gamma \delta} c_{\alpha}^\dagger c_{\beta}^\dagger c_{\gamma} c_{\delta} ,
\end{equation}
where $\epsilon_{\alpha}  $ represents the single-particle energy, $\overline{v}_{\alpha \beta \gamma \delta}= \langle \alpha \beta | V | \gamma \delta \rangle$~\cite{tbme,tbme2} are the two-body matrix elements (TBMEs) describing nucleon-nucleon interactions, and $ c_{\alpha}^\dagger,  c_{\alpha}$ are the creation and annihilation operators, respectively.

The kinetic energy contribution to the Hamiltonian is calculated as
\begin{equation}\label{eq:T}
\langle ab; JT | T_\epsilon | cd; JT \rangle = \delta_{ab} \delta_{cd} \frac{1 + \delta_{ab}}{1 - \delta_{cd}(-1)^{J+T}} (\epsilon_a + \epsilon_b),
\end{equation}
where $\quad a \leq b$.
The proton-neutron two-body matrix elements in the $M$-scheme~\cite{mscheme} are calculated using


\begin{eqnarray}
\overline{v}_{\alpha\beta\gamma\delta} &=& \sum_{JM,T} \left[ N_{ab}(JT) N_{cd}(JT) \right]^{-1} (j_am_\alpha j_bm_\beta | JM)\nonumber\\
&\times& (\frac{1}{2} m_{t\alpha} \frac{1}{2} m_{t\beta}|TM_T) (j_c m_\gamma j_d m_\delta | JM) \nonumber\\
&\times&  (\frac{1}{2} m_{t\gamma} \frac{1}{2} m_{t\delta} |TM_T) \langle ab; JT | V | cd; JT\rangle.
\end{eqnarray}
Here, $N_{ab}(JT)$ and $N_{cd}(JT)$ are normalization constants given as
\begin{eqnarray}
N_{ab}(JT) = \sqrt{\frac{1 - \delta_{ab} (-1)^{J+T}}{1 + \delta_{ab}}}.
\end{eqnarray}
This notation follows that of Ref.~\cite{suhonen2007nucleons}, where $a, b, c, d$ represent single-particle orbitals $|a\rangle = |n_\alpha, l_\alpha, j_\alpha \rangle$, and the Greek letters $\alpha, \beta, \gamma, \delta$ indicate the nucleon states with all quantum numbers, including isospin components. The terms inside the parenthesis are Clebsch-Gordan coefficients for coupling of angular momenta and isospins. \(\langle ab; J T \vert V \vert cd; J T \rangle\) are the two body matrix elements in $J$-scheme~\cite{suhonen2007nucleons}. 



\subsection{Hamiltonian encoding}\label{sec:encoding}
To simulate the nuclear systems on quantum computers, we need to encode the fermionic states $\ket{\psi}_f\in\mathcal{H}_{\rm nucl}$ to the qubit states $\ket{\psi}_q\in\mathcal{H}_{q}$, where $\mathcal{H}_{\rm nucl}$ and $\mathcal{H}_{q}$ stand for the fermionic and qubit Hilbert spaces, respectively. Under an encoding, the Hamiltonian in the second quantization form given in Eq.~\eqref{eq:H}, is transformed in terms of the qubit operator, which is a linear combination of Pauli strings, {\it i.e.}, $H=\sum x_iP_i$ where $P_i$ are the Pauli strings and $x_i$ are the coefficients. 

The creation and annihilation operators act on the fermionic states in the occupation number representations ($\ket{\psi}_f=\ket{f_{N-1}\ldots f_0}$ where $f_i=0$ and $1$ means vacant and occupied, respectively) in the following way
\begin{eqnarray}
    &c_i^{\dagger}&\ket{f_{N-1}\ldots f_i\ldots f_0}\nonumber\\
    &~&=\delta_{f_i,0}(-1)^{\sum_{k=0}^{i-1}f_k}\ket{f_{N-1}\ldots 1\oplus f_i\ldots f_0}
\end{eqnarray}
and 
\begin{eqnarray}
    &c_i&\ket{f_{N-1}\ldots f_i\ldots f_0}\nonumber\\
    &~&=\delta_{f_i,1}(-1)^{\sum_{k=0}^{i-1}f_k}\ket{f_{N-1}\ldots 1\oplus f_i\ldots f_0}
\end{eqnarray}
where $\oplus$ represents addition modulo 2 and factor $(-1)^{\sum_{k=0}^{i-1}f_k}$ accounts for the antisymmetrization of the wave function under the exchange operator. 

Several encodings have been proposed in the literature, for example, one-hot~\cite{JW:1928}, Bravyi-Kitaev~\cite{BK:2012}, parity, and Gray code~\cite{Siwach:PRC2021,dimatteo2021} encoding. The one-hot encoding is the most commonly used for mapping fermionic systems due to its simplicity. Under this encoding, the fermionic states in the occupation number representation are mapped directly to the qubit states $\ket{\psi}_q=\ket{q_{N-1}\ldots q_i\ldots q_0}$, such that
\begin{equation}
    \ket{f_{N-1}\ldots f_i\ldots f_0}\rightarrow\ket{q_{N-1}\ldots q_i\ldots q_0}.
\end{equation}
Therefore, in the one-hot encoding, the number of qubits $N$ required to map the fermionic system is the same as the number of degrees of freedom $N$ in the system. Correspondingly, the Jordan-Wigner (JW) transformation is used for expressing the Hamiltonian in terms of Pauli strings. The creation and annihilation operators under JW transformation take the following form
\begin{equation}
    c_i^\dagger = \left( \prod_{k=0}^{i-1} Z_k \right) \sigma_i^-, \quad 
c_i = \left( \prod_{k=0}^{i-1} Z_k \right) \sigma_i^+,
\end{equation}
where 
\begin{equation}
\sigma_i^\pm = \frac{1}{2}(X_i \pm \iota Y_i),
\end{equation}
$X_i$, $Y_i$, and $Z_i$ are the standard Pauli matrices acting on qubit $i$. The JW transformation inherently accounts for the anti-symmetrization of the wave function under the exchange operator. However, because of the large number of qubits required, it becomes quickly inefficient with an increase in the number of particles in the system. Furthermore, owing to the underlying symmetries, many irrelevant states exist in the qubit Hilbert space, making the simulations more error-prone. To overcome these challenges, in this work, we also explore the Gray code encoding in which the qubit requirement scales as $\lceil\log_2 N_b\rceil$ where $N_b$ is the number of basis states in $\mathcal{H_{\rm nucl}}$. In addition, the number of irrelevant states in the qubit Hilbert space is significantly reduced. For instance, let us consider a system of $m$ particles in $N$ number of orbitals. Total number of possible particle-conserving states are $K=\begin{pmatrix}N\\ m \end{pmatrix}$, however, not all of them might be contributing to the Hamiltonian matrix elements. In the case of one-hot encoding, we need $N$ qubits to map this system on a quantum computer. But in the case of GC, we need only $\lceil\log_2 Q\rceil$ qubits where $Q$ is the number of possible particle-conserving states contributing to the Hamiltonian. 

In the Gray code (GC) encoding, states are arranged such that only one qubit differs between adjacent states. This ordering has demonstrated clear advantages for simulations involving tri-diagonal Hamiltonians~\cite{dimatteo2021}. In Ref.~\cite{Siwach:PRC2021}, it is shown that the GC encoding exhibits superior performance in the presence of noise. While previous works have focused on applying GC encoding to bosonic~\cite{Sawaya:2020} and one-body fermionic systems~\cite{dimatteo2021,Siwach:PRC2021}, where antisymmetrization of wave functions is unnecessary, this study extends the approach to many-body fermionic systems. We present a detailed framework for transforming many-body fermionic operators into Pauli gates, ensuring the proper treatment of antisymmetrization required for fermionic wave functions.



We denote the fermionic states in GC encoding as $\ket{g_{Q-1}\ldots g_0}$ and directly map the $N_b$ fermionic basis states in the occupation number representation to these states with $Q=\lceil\log_2 N_b\rceil$ qubits
\begin{equation}
    \ket{f_{N-1}\ldots f_1f_0}\rightarrow\ket{g_{Q-1}\ldots g_1g_0}.
\end{equation}
To transform the Hamiltonian given in Eq.~\eqref{eq:H} in terms of Pauli strings, we need to transform the one-body $(c_i^{\dagger}c_j)$ and two-body $(c_i^{\dagger}c_j^{\dagger}c_kc_l)$ operators in terms of Pauli gates. The one-body operator $c_i^{\dagger}c_j$ can be written as
\begin{eqnarray}
     c_i^{\dagger}c_j&=&S^{ij}\ket{f_{N-1}\ldots f_i=1,\ldots f_j=0,\ldots f_0}\nonumber\\
     &~&\bra{f_{N-1}\ldots f_i=0,\ldots f_j=1,\ldots f_0},
\end{eqnarray}
where
\begin{equation}
    S^{ij}=\prod_{p={\rm min}(i,j)+1}^{{\rm max}(i,j)-1}(-1)^{f_p}
\end{equation}
accounts for the antisymmetrization of the wave function under exchange operation. In terms of GC, we have
\begin{eqnarray}
     c_i^{\dagger}c_j&=&S^{ij}\ket{g'_{Q-1}\ldots g'_1g'_0}\bra{g_{Q-1}\ldots g_1g_0},
\end{eqnarray}
where
$\ket{f_{N-1}\ldots f_i=1,\ldots f_j=0,\ldots f_0}$ is mapped to $\ket{g'_{Q-1}\ldots g'_1g'_0}$ and $\ket{f_{N-1}\ldots f_i=0,\ldots f_j=1,\ldots f_0}$ is mapped to $\ket{g_{Q-1}\ldots g_1g_0}$.

Similarly, for the two-body operators $c_i^{\dagger}c_j^{\dagger}c_kc_l$, the non-zero matrix elements are of the form
\begin{widetext}
\begin{eqnarray}
     c_i^{\dagger}c_j^{\dagger}c_kc_l=S^{ij}S^{jk}&S^{kl}&\ket{f_{N-1}\ldots f_i=1\ldots f_j=1\ldots f_k=0\ldots f_l=0\ldots f_0}\nonumber\\
     &\times&\bra{f_{N-1}\ldots f_i=0\ldots f_j=0\ldots f_k=1\ldots f_l=1\ldots f_0}\nonumber\\
     =S^{ij}S^{jk}&S^{kl}&\ket{g'_{Q-1}\ldots g'_1g'_0}\bra{g_{Q-1}\ldots g_1g_0}.
\end{eqnarray}
\end{widetext}
The higher-order operators can be transformed in the same way.

To transform the operators $\ket{g'_{Q-1}\ldots g'_1g'_0}\bra{g_{Q-1}\ldots g_1g_0}$, we can consider them in the following form
\begin{eqnarray}
    \ket{g'_{Q-1}\ldots g'_1g'_0}&\bra{g_{Q-1}\ldots g_1g_0}&\nonumber\\
    =&\ket{g'_{Q-1}}\bra{g_{Q-1}}&\otimes\ldots\otimes\ket{g'_0}\bra{g_0}.
\end{eqnarray}
The operators of the form $\ket{g'_Q}\bra{g_Q}$ can be transformed in terms of Pauli matrices by using the following expressions.
\begin{eqnarray}
    \ket{0}\bra{1}&=&\sigma^{+}=\frac{1}{2}\left(X+iY\right),\\
    \ket{1}\bra{0}&=&\sigma^{-}=\frac{1}{2}\left(X-iY\right),\\
    \ket{0}\bra{0}&=&P^{(0)}=\frac{1}{2}\left(I+Z\right), \text{and} \\
    \ket{1}\bra{1}&=&P^{(1)}=\frac{1}{2}\left(I-Z\right).
\end{eqnarray}
To illustrate the transformations under Gray code encoding, we provide an example in appendix~\ref{app:encoding}.
\subsection{Particle-conserving ansatz}\label{sec:ansatz}
The structure of the ansatz is a critical factor in the success of VQE. An ansatz that spans irrelevant subspaces of the Hilbert space not only increases the number of parameters to optimize, but also introduces additional errors. To address these challenges, several ansatz designs have been proposed that reduce the number of quantum gates while preserving essential symmetries, such as particle number and spin symmetry. In the nuclear shell model, where the number of valence particles outside the core is fixed, ansatz must inherently conserve particle number. In a previous work~\cite{siwach:prc2022}, a strategy was proposed to prepare a quantum state by rotating the qubits using controlled rotation-$Y$ gates with an angle depending on the coefficients of the states and two-qubit CNOT gates. Building on this framework, we propose a simplified particle-conserving ansatz that reduces gate count and circuit depth while maintaining particle-conserving properties. Furthermore, our approach can be extended to design universal circuits, similar to the ones proposed in Ref.~\cite{Arrazola:quantum2022} but with a smaller number of gates and shallower circuits, capable of preparing a wide range of particle-conserving quantum states efficiently.

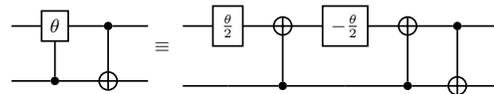
\begin{figure}[h!]
\begin{tikzpicture}
\node[scale=0.8] {
\begin{quantikz}
\qw&\gate{\theta}&\ctrl{1}&\qw\\
\qw&\ctrl{-1}&\targ{}&\qw
\end{quantikz}
$\equiv$
\begin{quantikz}
\qw&\gate{\frac{\theta}{2}}&\targ{}&\gate{-\frac{\theta}{2}}&\targ{}&\ctrl{1}&\qw\\
\qw&\qw&\ctrl{-1}&\qw&\ctrl{-1}&\targ{}&\qw
\end{quantikz}
};
\end{tikzpicture}
\caption{Quantum circuit to prepare superposition of states differing by single excitations. Here gate $\theta$ represents the rotation-$Y$ gate.}\label{fig:se}
\end{figure}

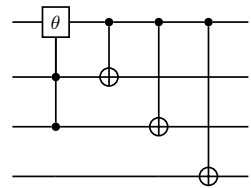
\begin{figure}
\begin{tikzpicture}
\node[scale=0.8] {
\begin{quantikz}
\qw&\gate{\theta}&\ctrl{1}&\ctrl{2}&\ctrl{3}&\qw\\
\qw&\ctrl{-1}&\targ{}&\qw&\qw&\qw\\
\qw&\ctrl{-2}&\qw&\targ{}&\qw&\qw\\
\qw&\qw&\qw&\qw&\targ{}&\qw
\end{quantikz}
};
\end{tikzpicture}

\caption{Quantum circuit to prepare superposition of states differing by double excitations.}\label{fig:de}
\end{figure}

\begin{figure*}
\begin{tikzpicture}
\node[scale=0.9] {
\begin{quantikz}
\qw&\gate{\theta}&\ctrl{1}&\ctrl{2}&\ctrl{3}&\qw\\
\qw&\ctrl{-1}&\targ{}&\qw&\qw&\qw\\
\qw&\ctrl{-2}&\qw&\targ{}&\qw&\qw\\
\qw&\qw&\qw&\qw&\targ{}&\qw\\
\end{quantikz}
$\equiv$
\begin{quantikz}
\qw&\targ{}&\gate{-\frac{\theta}{4}}&\targ{}&\gate{\frac{\theta}{4}}&\targ{}&\gate{\frac{\theta}{4}}&\targ{}&\gate{-\frac{\theta}{4}}&\targ{}&\gate{-\frac{\theta}{4}}&\targ{}&\gate{\frac{\theta}{4}}&\ctrl{1}&\ctrl{2}&\ctrl{3}&\qw\\
\qw&\qw&\qw&\qw&\targ{}&\ctrl{-1}&\qw&\ctrl{-1}&\targ{}&\ctrl{-1}&\qw&\ctrl{-1}&\qw&\targ{}&\qw&\qw&\qw\\
\qw&\ctrl{-2}&\qw&\ctrl{-2}&\ctrl{-1}&\qw&\qw&\qw&\ctrl{-1}&\qw&\qw&\qw&\qw&\qw&\targ{}&\qw&\qw\\
\qw&\qw&\qw&\qw&\qw&\qw&\qw&\qw&\qw&\qw&\qw&\qw&\qw&\qw&\qw&\targ{}&\qw
\end{quantikz}
};
\end{tikzpicture}

\caption{The decomposition of double excitation circuit in terms of single- and two-qubit gates.}\label{fig:decomp}
\end{figure*}
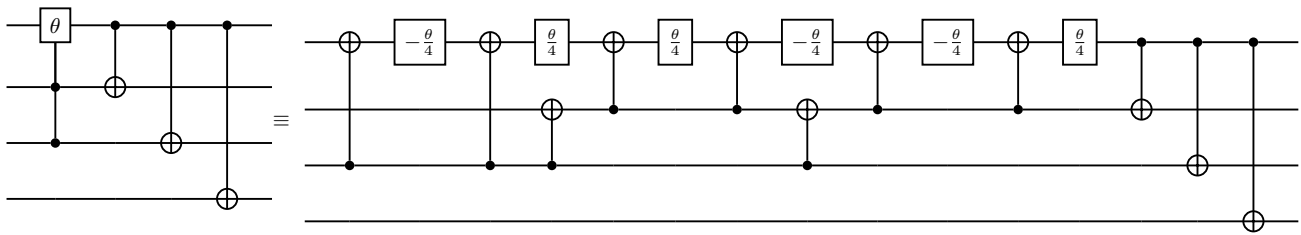

A superposition of states in the following form
\begin{eqnarray}
    \ket{\psi}&=&c_1\ket{f_{N-1}\ldots 0_j\ldots 1_{j'}\ldots f_0}\nonumber\\
    &+&c_2\ket{f_{N-1}\ldots 1_j\ldots 0_{j'}\ldots f_0}
\end{eqnarray}
can be prepared by applying a single-excitation circuit given in Figure~\ref{fig:se} on qubit $j$ and $j'$. A similar circuit is given in Ref.~\cite{Arrazola:quantum2022} with four CNOTs and two single-qubit rotation-$Y$ gates. We realize that this operation can be achieved with a reduced CNOT count by one. A similar state with double excitations can be prepared with the circuit given in Figure~\ref{fig:de}, which can be further decomposed in single- and two-qubit CNOT gates as shown in Figure~\ref{fig:decomp}. For example, to create a superposition of $\ket{000011}$ and $\ket{001100}$, {\it i.e.,} $c_1\ket{000011} +c_2\ket{001100}$ we can apply the circuit given in Figure~\ref{fig:de} on the first, second, third and fourth qubits.

To apply the single-excitation circuit in a multi-qubit case, it is applied with control operation on the qubits which are 1 in both the states simultaneously. For example, to create a superposition of $\ket{0011}$ and $\ket{0101}$, the single excitation circuit is applied to qubits first and second with control operation on the zeroth qubit, which is 1 in both states simultaneously. Similarly, in the case of double excitations, the circuit is controlled by qubits which are 1 in both the states but these controlled operations are required only for the rotation-$Y$ gate not the CNOT gates unlike mentioned in Ref.~\cite{Arrazola:quantum2022}. Therefore, we achieve a circuit with significantly reduced number of gates and reduced circuit depth. For example, the double excitation circuit can be built only with 6 single-qubit and 11 CNOT gates compared to the 14 single-qubit and 14 CNOT gates given in Ref.~\cite{Arrazola:quantum2022}. The difference in gate counts will increase further with an increase in the number of qubits. Though we mention only single- and double-excitation operations, our technique is applicable to higher excitations as well.

To design the ansatz, $\theta$s are kept as parameters. To prepare a particle-conserving state with fixed coefficients $c_i$s, $\theta$s can be replaced with
\begin{equation}
    \frac{\theta_i}{2}=\cos^{-1}\left(\frac{c_i}{\sqrt{1-\sum_{j<i}c_j^2}}\right)
\end{equation}
We provide an illustrative example in appendix~\ref{app:ansatz} to design an ansatz using the above mentioned procedure.

\section{Quantum algorithms}\label{sec:sec2}
In this section, various variational methods to study nuclear systems are explored with an aim to determine both ground state and excited state energies efficiently. 

\subsection{VQE}
The variational quantum eigensolver (VQE)~\cite{VQE_nature1,VQE_nature2, Siwach:PRC2021} algorithm is designed to find the lowest eigenvalue of a Hamiltonian, leveraging the variational principle of quantum mechanics. This principle states that the expectation value of the Hamiltonian with respect to any state is always greater than or equal to its minimum eigenvalue. VQE is a hybrid quantum-classical algorithm. On the quantum device, an ansatz (a trial wave function) $\ket{\psi(\theta)}$ is prepared by applying quantum gates to qubits, and the expectation value of the Hamiltonian is measured. 
\begin{equation}\label{eq:vqeCost}
    E(\theta)=\bra{\psi(\theta)}H\ket{\psi(\theta)}
\end{equation}
This measured value {\it i.e.}, energy $E(\theta)$ is then minimized using a classical optimizer by tuning the parameter $\theta$ of the ansatz. Due to its use of shallower circuits compared to the more resource-intensive quantum phase estimation algorithm, VQE is particularly well-suited to near-term quantum devices. However, the algorithm's accuracy is highly dependent on the choice and structure of the ansatz, which plays a critical role in achieving reliable results.

In this work, we utilize the particle-conserving form of ansatz given in section~\ref{sec:ansatz} for the Jordan-Wigner transformation. For Gray code encoding, we use the hardware-efficient variational ansatz, which is developed for up to $3$ qubits in Refs.~\cite{Siwach:PRC2021, dimatteo2021}. We expand this approach to construct the ansatz for $4$ qubits [Fig.~\ref{fig:adapt0+ligcvqe}]. 

\subsection{ADAPT-VQE}\label{sec:qubit_adapt}
The form of the ansatz is fixed in the VQE algorithm which might lead to deeper circuits and a large number of parameters to be tuned. An improvement in VQE is suggested in the form of adaptive derivative assembled pseudo-trotter (ADAPT) VQE~\cite{Grimsley:nc2019,adapt,Romero:2022}. In this algorithm, the form of the ansatz is built dynamically from a set of operators called ``operator pool". The operators $\hat{A}_i$ from the operator pool are applied iteratively on a reference state $\ket{\psi^{\rm ref}}$ such that the ansatz after the $n^{th}$ iteration takes the following form. 
\begin{equation}
    \ket{\psi^{\rm ADAPT}(\vec{\theta})}={\rm e}^{\theta_n \hat{A}_n}\ldots{\rm e}^{\theta_2 \hat{A}_2}{\rm e}^{\theta_1 \hat{A}_1}\ket{\psi^{\rm ref}},
\end{equation}
where $\vec{\theta}=\{\theta_1,\theta_2,\ldots,\theta_n\}$.
At each iteration, the commutator of operators with the Hamiltonian is measured and the gradient of the energy is calculated with respect to the corresponding parameter as follows
\begin{equation}
    \frac{\partial}{\partial \theta_i}\braket{E}=\bra{\psi(\vec{\theta})}\left[\hat{H},\hat{A}_i\right]\ket{\psi(\vec{\theta})}.
\end{equation}
Since the measurements for all operators are independent, this step can be executed in parallel. The operator with the largest gradient is then selected and appended to the leftmost position of the ansatz in the next iteration. After updating the ansatz, the VQE algorithm is used to re-optimize its parameters. This iterative process continues, with the ansatz growing incrementally, until the norm of the gradient falls below a predefined threshold, indicating convergence. In our case, we set the gradient threshold to $10^{-5}$ relative to the ground state energy. 

ADAPT-VQE requires a greater number of measurements compared to standard VQE. The number of measurements scales with the size of the operator pool and the number of Pauli strings in the Hamiltonian. In its original form, known as fermionic-ADAPT~\cite{Grimsley:nc2019}, the operator pool is constructed using fermionic operators, specifically single- and double-excitation operators. While these operators result in a parameter-efficient ansatz, they require a large number of CNOT gates, making the approach gate-inefficient. To address this limitation, a more gate-efficient scheme called qubit-ADAPT-VQE was proposed in Ref.~\cite{Tang:PRXQuantum2021}. Qubit-ADAPT-VQE constructs the operator pool directly in the qubit space, significantly reducing the CNOT gate count.

In qubit-ADAPT-VQE, the operator pool is constructed by decomposing fermionic operators into their corresponding Pauli strings obtained from JW transformation. Each individual Pauli string is then selected as an element of the operator pool such as $\hat{A}_i=i\prod_ip_i, ~p_i\in\{X,Y,Z\}.$ This reduced pool is called \emph{qubit pool} and contains only strings with odd numbers of $Y$'s.

For $n$ number of qubits, the minimal pool $\{V_j\}_n$~\cite{Tang:PRXQuantum2021} for $j=1,\ldots,2n-2$, that generates a complete basis for an arbitrary state can be constructed recursively as
\begin{equation}
    \{V_j\}_n=\{Z_n\{V_k\}_{n-1},iY_n,iY_{n-1}\},
\end{equation}
with only $2n-2$ operators which begins with $n=2$ qubits
\begin{equation}
    \{V_j\}_2=\{iZ_2Y_1,iY_2\}.
\end{equation}
This $\{V_j\}$ can be mapped to another set of minimal complete pools $\{G_j\}$, which contains all two-qubit operators of the form $i Z_{n}Y_{n-1}$, and single qubit operators of form $iY_{n-1}$. This pool also contains $2n-2$ operators but requires fewer gates compared to the set $\{V_j\}$. We use the $\{G_j\}$ operator pool in our calculations for both JW and GC encoding.
As an example for $n=3$, the sets are 
\begin{align}
    \{V_j\}_3& = \{iZ_3Z_2Y_1, iZ_3Y_2, iY_3, iY_2\}, \text{and}\\
    \{G_j\}_3& =\{iZ_2Y_1, iZ_3Y_2, iY_2, iY_3\}.
\end{align}

\subsection{Excited states calculation}\label{sec:vqd}
The algorithms VQE and ADAPT-VQE are designed to calculate the lowest eigenvalue of a Hamiltonian. Several modifications have been suggested to enable VQE to find the excited states. In the present work, we utilize the variational quantum deflation (VQD) algorithm~\cite{Higgott:quantum2019, vqd2} to calculate the excited state eigenvalues of the nuclear shell model Hamiltonian. It extends the VQE to calculate the $k$-th eigenvalue of the Hamiltonian by optimizing the parameters $\theta_k$ to minimize the cost function $F(\theta_k)$ given by
\begin{equation}
    F(\theta_k)=\bra{\psi(\theta_k)}H\ket{\psi(\theta_k)}+\sum_{i=0}^{k-1}\beta_i\braket{\psi(\theta_k)|\psi(\theta_i)}
\end{equation}
Here, the overlap term ensures that the $k$-th state $\ket{\psi(\theta_k)}$ is orthogonal to the previously computed states $\ket{\psi(\theta_0)},\ldots,\ket{\psi(\theta_{k-1})}$. The weights $\beta_i$ of the overlaps are chosen such that the $F(\theta_k)$ minimizes the energy of the $k$-th excited state efficiently. Starting with the ground state, the excited state energies are calculated iteratively. 

In this work, we integrate ADAPT-VQE with VQD to compute the excited states of the nuclear shell model Hamiltonian. The ground state ansatz is constructed using ADAPT-VQE, which is subsequently employed to compute excited states with VQD. Additionally, to highlight the advantages and differences between the two approaches, we also perform excited-state calculations using VQD with VQE.



\section{Results}\label{sec:Results}
In this section, we present the results of our quantum calculations for the full energy spectrum, including both ground and excited states for the $^{38}$Ar and $^6$Li nuclei using the JW and GC schemes. The exact energy values correspond to the results obtained from classical computation and serve as benchmarks for comparison with the results obtained using qubit-ADAPT-VQE, VQE, and VQD methods in both noiseless and noisy conditions.

\subsection{Numerical settings}
We perform the quantum simulations to calculate the full energy spectrum of nuclei $^{38}$Ar and $^6$Li. For the case of $^{38}$Ar, we model the nucleus as a $^{28}$Si core with two holes in $sd$-shell. The valence nucleons are treated as single-particle states within the $sd$-shell model space ($s_{1/2}$ and $d_{3/2}$) and utilize USDB~\cite{sd} nucleon-nucleon interaction. In the $m$-scheme, the mapping of the single-particle orbitals requires $N = 6$ qubits for six proton orbitals as shown in Table~\ref{tab:38Arbasis}.
\begin{table}[h!]
    \centering
    \caption{Mapping of single-particle orbitals in $sd$-shell ($s_{1/2}$ and $d_{3/2}$) on qubits in $m$-scheme with $t_z=-1/2$ for all cases.}
    \label{tab:38Arbasis}
    \setlength{\tabcolsep}{8pt}
    \renewcommand{\arraystretch}{1.6}
    \begin{tabular}{|c|c|c|c|r|}
    \hline
    Qubit & $n$ & $l$ & $2j$ & $2j_z$ \\
    \hline
    $0$ &  $0$ & $2$ & $3$ & $-1$ \\
    $1$ &  $1$ & $0$ & $1$ & $-1$ \\
    $2$ &  $0$ & $2$ & $3$ & $-3$ \\
    $3$ &  $0$ & $2$ & $3$ & $ 1$ \\
    $4$ &  $1$ & $0$ & $1$ & $ 1$ \\
    $5$ &  $0$ & $2$ & $3$ & $ 3$ \\
\hline
    \end{tabular}
\end{table}

For the case of $^6$Li, we model the nucleus as a $^4$He core with two valence nucleons {\it i.e.}, one proton and one neutron. The valence nucleons are treated as single-particle states within the $p$-shell model space and utilize Cohen-Kurath~\cite{p} nucleon-nucleon interaction. In the $J$-scheme, the basis states are defined as $|i\rangle = |n = 0, l = 1, j, t_z\rangle$, where $n$ and $l$ are the radial and orbital angular momentum quantum numbers, $j = \frac{1}{2}, \frac{3}{2}$ is the total spin, and $t_z$ is the isospin projection $(t_z = \frac{1}{2}$ for neutrons and $t_z = -\frac{1}{2}$ for protons). This scheme requires $N = 4$ qubits, as each orbital is mapped to one qubit (see Table~\ref{tab:6LibasisJscheme}).

\begin{table}[h!]
    \centering
    \caption{Mapping of single-particle orbitals in $p$-shell on qubits in $J$-scheme.}
    \label{tab:6LibasisJscheme}
    \setlength{\tabcolsep}{6pt}
    \renewcommand{\arraystretch}{1.6}
    \begin{tabular}{|c|c|r|}
    \hline
    Qubit & $2j$ & $2t_z$\\
    \hline
    0 & $1$ & $-1$ \\
    1 & $3$ & $-1$ \\
    2 & $1$ & $+1$ \\
    3 & $3$ & $+1$ \\
\hline
    \end{tabular}   
\end{table}

\begin{table}[h!]
    \centering
    \caption{Mapping of single-particle orbitals in $p$-shell on qubits in $m$-scheme.}
    \label{tab:6Libasis}
    \setlength{\tabcolsep}{8pt}
    \renewcommand{\arraystretch}{1.6}
    \begin{tabular}{|c|c|c|c|r|r|}
    \hline
    Qubit & $n$ & $l$ & $2j$ & $2j_z$ & $2t_z$\\
    \hline
    $0$ &  $0$ & $1$ & $1$ & $-1$ & $1$ \\
    $1$ &  $0$ & $1$ & $3$ & $-1$ & $1$ \\
    $2$ &  $0$ & $1$ & $3$ & $-3$ & $1$ \\
    $3$ &  $0$ & $1$ & $1$ & $1$ & $1$ \\
    $4$ &  $0$ & $1$ & $3$ & $1$ & $1$ \\
    $5$ &  $0$ & $1$ & $3$ & $3$ & $1$ \\
    $6$ &  $0$ & $1$ & $1$ & $-1$ & $-1$ \\
    $7$ &  $0$ & $1$ & $3$ & $-1$ & $-1$ \\
    $8$ &  $0$ & $1$ & $3$ & $-3$ & $-1$ \\
    $9$ &  $0$ & $1$ & $1$ & $1$ & $-1$ \\
    $10$ & $0$ & $1$ & $3$ & $1$ & $-1$ \\
    $11$ & $0$ & $1$ & $3$ & $3$ & $-1$ \\
\hline
    \end{tabular}
\end{table}
The $m$-scheme representation provides a more detailed description of the states. Here, the basis states are written as $|i\rangle = |n = 0, l = 1, j, j_z, t_z\rangle$, where $j_z$ is the projection of the total spin $j$. This scheme includes all possible $j_z$ projections, requiring $N = 12$ qubits for six proton orbitals and six neutron orbitals as shown in Table~\ref{tab:6Libasis}.

\subsection{$^{38}$Ar}

\begin{table*}
\centering
\caption{Energies of lower-lying levels in $^{38}$Ar calculated in JW scheme.}
\label{tab:Ar_jw_table_0}
\setlength{\tabcolsep}{8pt}
\renewcommand{\arraystretch}{1.6}
\begin{tabular}{|c|c|l|c|l|}
\hline
 $J^{\pi}$ &\textbf{Exact} & \textbf{Noise level} & \textbf{ADAPT-VQE } & \textbf{VQE} \\ \hline
 \multirow{3}{*}{$0^+$} & \multirow{3}{*}{$-152.677$} & Without Noise    &$-152.3999 \pm 0.8798$& $-151.7937 \pm 0.4477$  \\ \cline{3-5}
& & With Noise & $-141.6124 \pm 7.2305$ & $-139.8726 \pm 9.4223$\\ \cline{3-5}
\hline
\multirow{3}{*}{$2^+$} & \multirow{3}{*}{$-151.093$} & Without Noise   & $-150.8583 \pm 1.1249$  &  $-151.0934 \pm 0.0018$\\ \cline{3-5}
& & With Noise & $-142.1343 \pm 4.5646$  & $-149.6657 \pm 0.5218$ \\ \cline{3-5}
\hline
\multirow{3}{*}{$2^+$}& \multirow{3}{*}{$-149.700$}& Without Noise      & $-149.7034 \pm 0.0001$    & $-149.7003$ \\ \cline{3-5}
& & With Noise     &   $-149.5180 \pm 0.5566$      &  $-149.7002 \pm 0.0002$    \\ \cline{3-5}
\hline
\multirow{3}{*}{$0^+$} & \multirow{3}{*}{$-149.225$} & Without Noise   &  $-140.6414 \pm  14.4897$ & $-137.4674 \pm 18.7408$   \\ \cline{3-5}
& & With Noise & $-129.7738 \pm 21.0019$&$-110.8731 \pm 30.5988$  \\ \cline{3-5}
\hline
\multirow{3}{*}{$1^+$} & \multirow{3}{*}{$-149.104$} & Without Noise   &$-148.5788 \pm 4.2723$ & $-148.8602 \pm 0.6486$   \\ \cline{3-5}
& & With Noise &$-141.8163 \pm 2.0340$ &$-136.1554 \pm 4.5441$  \\ \cline{3-5}
\hline
\end{tabular}
\end{table*}

\begin{table}
\centering
\caption{Comparison of number of parameters P($\theta$) and gate count in the 6 qubit ansätze (Appendix~\ref{app:result_ansatz}) used for low lying states of $^{38}$Ar in JW scheme.}
\label{tab:gate_ar_jw}
\renewcommand{\arraystretch}{2}
\begin{tabular}{|c|c|c|c|c|c|c|}
\hline
\textbf{$J^{\pi}$} & \multicolumn{3}{c|}{\textbf{ADAPT-VQE}} & \multicolumn{3}{c|}{\textbf{VQE}} \\
\cline{2-7}
 & P($\theta$) & 1Q & 2Q & P($\theta$) & 1Q & 2Q \\
\hline
$0^+$ & 13 & 39 & 14 & 4 & 57 & 35 \\
\hline
$1^+$ & 9  & 31 & 10 & 2 & 5  & 6  \\
\hline
$2^+$ & 7  & 31 & 10 & 1 & 3  & 1  \\
\hline
\end{tabular} 
\end{table}
\begin{table*}
\centering
\caption{Energies of lower-lying levels in $^{38}$Ar calculated in GC scheme.}
\label{tab:Ar_gc_table_0}
\setlength{\tabcolsep}{8pt}
\renewcommand{\arraystretch}{1.6}
\begin{tabular}{|c|c|l|l|l|}
\hline
 $J^{\pi}$ &\textbf{Exact} & \textbf{Noise level} & \textbf{ADAPT-VQE } & \textbf{VQE} \\ \hline
 \multirow{3}{*}{$0^+$} & \multirow{3}{*}{$-152.677$} & Without Noise    & $-152.6669 \pm  0.6646$    & $ -152.1847 \pm 0.6602$   \\ \cline{3-5}
& & With Noise & $-148.5119 \pm 1.4785$      & $-148.1243 \pm 1.8329$ \\ \cline{3-5}
\hline
\multirow{3}{*}{$2^+$} & \multirow{3}{*}{$-151.093$} & Without Noise   & $-151.0928 \pm 0.0019$ & $-151.0934 \pm 0.0025$   \\ \cline{3-5}
& & With Noise & $ -151.0295 \pm 0.0033$  & $-151.0298\pm  0.0024$ \\ \cline{3-5}
\hline
\multirow{3}{*}{$2^+$}& \multirow{3}{*}{$-149.700$}& Without Noise      & $-149.7002 $      & $-149.7003 $       \\ \cline{3-5}
& & With Noise     &   $-149.7002 $  & $-149.7002$    \\ \cline{3-5}
\hline
\multirow{3}{*}{$0^+$} & \multirow{3}{*}{$-149.225$} & Without Noise   &  $ -149.1965 \pm 0.8576$ &   $ -148.6076 \pm 0.4396$  \\ \cline{3-5}
& & With Noise &  $-144.4054 \pm 2.4409 $  & $ -141.3890 \pm 3.5595$ \\ \cline{3-5}
\hline
\multirow{3}{*}{$1^+$} & \multirow{3}{*}{$-149.104$} & Without Noise   &  $-149.0768 \pm 0.4187$   &   $-148.9775 \pm 0.5349$  \\ \cline{3-5}
& & With Noise & $-148.9629 \pm 0.4208 $   &  $-148.9016 \pm 0.6109$ \\ \cline{3-5}
\hline
\end{tabular}
\end{table*}

\begin{table}
\centering
\caption{Comparison of number of parameters P($\theta$) and gate count in the ansätze (Appendix~\ref{app:result_ansatz}) used for low lying states of $^{38}$Ar in GC scheme.}
\label{tab:gate_ar_gc}
\renewcommand{\arraystretch}{2}
\begin{tabular}{|c|c|c|c|c|c|c|}
\hline
\textbf{$J^{\pi}$} & \multicolumn{3}{c|}{\textbf{ADAPT-VQE}} & \multicolumn{3}{c|}{\textbf{VQE}} \\
\cline{2-7}
 & P($\theta$) & 1Q & 2Q & P($\theta$) & 1Q & 2Q \\
\hline
$0^+$ & 4 & 14 & 4 & 7 & 7 & 4 \\
\hline
$1^+$ & 3  & 9 & 2 & 3 & 3  & 1  \\
\hline
$2^+$ & 1  & 2 & 0 & 1 & 1  & 0  \\
\hline
\end{tabular}
\end{table}

\subsubsection{JW scheme}
The energy spectrum of $^{38}$Ar calculated with quantum simulations under one-hot encoding and JW transformation are shown in Table~\ref{tab:Ar_jw_table_0}. We compare the exact energy values obtained from classical computation with the ones obtained from quantum simulations under noiseless and noisy conditions.  The energies of lowest-lying states with spin $0^+$, $1^+$, and $2^+$ are calculated using both qubit-ADAPT-VQE and standard VQE. The dynamically constructed ansätze for qubit-ADAPT-VQE using the qubit operator pool are given in appendix~\ref{app:ansatz_ADAPT}. For VQE, a particle conserving ansatz, as described in section~\ref{sec:ansatz}, is used for ground state calculations. The specific ansätze for $0^+$, $1^+$, and $2^+$ states are given in appendix~\ref{app:ansatz_VQE}. In both calculations, the simultaneous perturbation stochastic approximation (SPSA) optimizer~\cite{spsa} was used, with $10000$ shots per run, and the results given are the median of $100$ runs with error computed as median absolute deviation (MAD) to ensure statistical accuracy.

To better understand the performance differences between the two methods, Table~\ref{tab:gate_ar_jw} shows a comparison of the number of parameters and gate counts for the ansätze used. The result clearly shows that accuracy in the quantum simulation is strongly related to the depth and complexity of the ansatz. The single-qubit gates ($1$Q) have relatively less impact on the overall fidelity, while the number of parameters P($\theta$) and two-qubit gates ($2$Q) influence the performance significantly. 

For the $1^+$ and $2^+$ states, VQE gives more accurate results. This aligns with the fact that particle conserving ansätze for these states uses fewer parameters and gates than ADAPT-VQE. However, for the $0^+$ state, qubit-ADAPT-VQE outperforms VQE due to its dynamically constructed ansatz, which results in a shallower and more efficient circuit compared to the relatively deeper VQE ansatz for this state.

For states with higher energies (second $0^+$, and $2^+$), the variational quantum deflation (VQD) algorithm is used, as explained in section~\ref{sec:vqd}. Similar to the lowest-lying state, VQD combined with qubit-ADAPT-VQE gives better results for the $0^+$ state.  However, for the $2^+$ state, VQD with VQE ansatz produces comparable results due to its smaller circuit depth. To maintain the similar order of accuracy in the results, $10^{14}$ shots are taken for each run, and final results were taken as the median of $100$ runs with MAD error.



\subsubsection{GC scheme}
Results for $^{38}$Ar using GC encoding are given in Table~\ref{tab:Ar_gc_table_0}. The reduced number of qubits and gates in GC encoding significantly improves the accuracy of computed energies compared to JW, which uses 6 qubits for all $J^{\pi}$ states. In contrast, GC encoding requires only 3 qubits for the $0^+$ state, 2 qubits for the $1^+$ state, and just 1 qubit for the $2^+$ state. The calculations for the lowest energy states for spins $0^+$, $1^+$, and $2^+$ are performed using VQE with hardware-efficient variational ansätze consisting of layers of parametrized $Y$ rotations separated by layers of entangling gates, as shown in appendix~\ref{app:ansatz_VQE}.

The final ansätze constructed using qubit-ADAPT-VQE for $0^+$, $1^+$, and $2^+$ are given in appendix~\ref{app:ansatz_ADAPT}. Similar to JW encoding, qubit-ADAPT-VQE outperformed VQE for $0^+$ and $1^+$ state due to its compact dynamically constructed ansätze. However, the accuracy is significantly improved in GC encoding, as the ansätze (Table~\ref{tab:gate_ar_gc}) requires significantly fewer computational resources compared to JW (Table~\ref{tab:gate_ar_jw}).

For the $2^+$ state, results from both qubit-ADAPT-VQE and VQE are nearly identical due to the comparable parameters and gate counts, as seen in Table~\ref{tab:gate_ar_gc}. The difference becomes more pronounced when fewer shots are used, though in our simulations, the number of shots, runs, and optimizer settings are kept identical to show a direct comparison with JW. 

For excited states (second $0^+$ and $2^+$), GC encoding continues to enhance accuracy owing to the reduced qubit requirement and overall circuit depth. As with the ground states, ADAPT-VQE gives better results for the $0^+$ state, while for $2^+$, both give comparable results. 

\begin{figure}
   \centering
    \includegraphics[width = \linewidth]{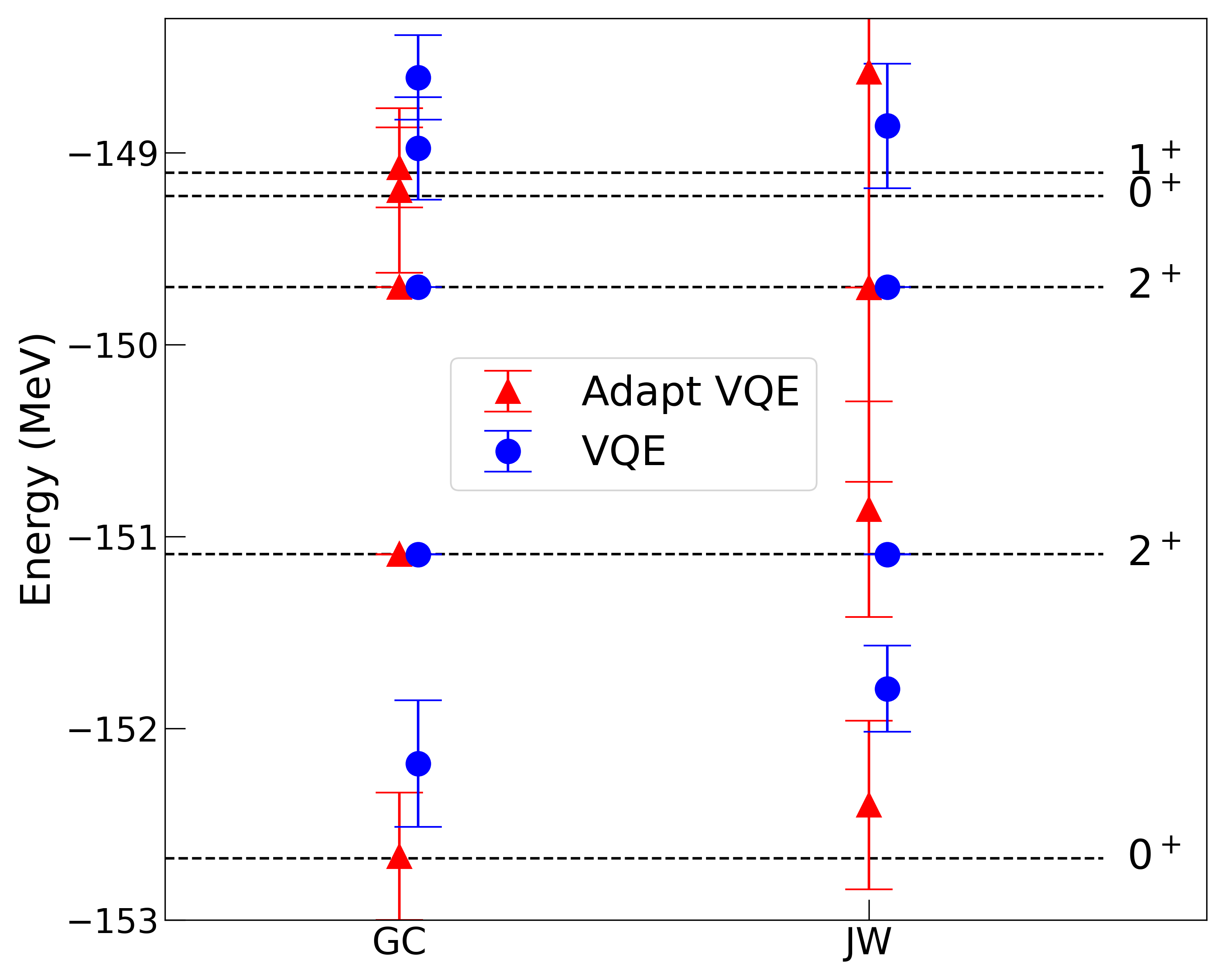}
    \caption{Level scheme of $^{38}$Ar calculated using ADAPT-VQE and VQE with GC and JW schemes (without noise). The black dashed lines shows the exact energy value for state ($J^{\pi}$).} 
    \label{fig:gc_ar_jw}
\end{figure}

The overall comparison between GC and JW results for $^{38}$Ar is shown in Fig.~\ref{fig:gc_ar_jw}. It shows the level scheme of $^{38}$Ar computed with ADAPT-VQE and VQE using both encoding schemes. The exact energy value for each state ($J^{\pi}$) are indicated by black dashed lines. ADAPT-VQE provides significantly better accuracy for larger qubit states, while for states with lesser qubits, both give almost comparable results, with VQE performing slightly better. Additionally, GC encoding consistently outperforms JW across all states and methods. Even the circuits under GC are less sensitive to noise, highlighting that GC not only improves precision but also reduces the impact of quantum hardware noise by requiring shallower and simpler circuits.



\subsection{$^6$Li}

\begin{table*}
\centering
\caption{Energies of lower-lying levels in $^6$Li calculated in GC scheme.}
\label{tab:Li_gc_table_0}
\setlength{\tabcolsep}{8pt}
\renewcommand{\arraystretch}{1.6}
\begin{tabular}{|c|c|l|l|l|}
\hline
 $J^{\pi}$ &\textbf{Exact} & \textbf{Noise level} & \textbf{ADAPT-VQE } & \textbf{VQE} \\ \hline
 \multirow{3}{*}{$1^+$} & \multirow{3}{*}{$-5.433$} & Without Noise   & $-5.4254 \pm 0.0459$   & $-5.3757 \pm  0.1555$    \\ \cline{3-5}
& & With Noise & $ -5.1101 \pm 1.0181$& $-4.9938 \pm 1.3217$\\ \cline{3-5}
\hline
\multirow{3}{*}{$3^+$} & \multirow{3}{*}{$-5.009$} & Without Noise   &  $-5.0089$ & $-5.0099$   \\ \cline{3-5}
& & With Noise & $-5.0089 \pm 0.0001$ & $-5.0089$ \\ \cline{3-5}
\hline
\multirow{3}{*}{$0^+$}& \multirow{3}{*}{$-3.910$}& Without Noise      & $-3.6782 \pm 0.5792$      &  $-3.0626 \pm 0.6780$   \\ \cline{3-5}
& & With Noise     &  $ -3.1198 \pm 1.3085$  &  $-2.8558 \pm 1.5756$ \\ \cline{3-5}
\hline
\multirow{3}{*}{$1^+$} & \multirow{3}{*}{$-1.273$} & Without Noise   &  $-1.2331 \pm 0.9776$ & $ -1.1941 \pm  0.0815$    \\ \cline{3-5}
& & With Noise & $-1.1814 \pm 0.8276$ & $-1.1560 \pm 1.0519$ \\ \cline{3-5}
\hline
\multirow{3}{*}{$2^+$} & \multirow{3}{*}{$-0.510$} & Without Noise   & $-0.5108 \pm 0.0018$ &  $-0.5097 \pm 0.0002$    \\ \cline{3-5}
& & With Noise & $-0.4925 \pm 0.0789$   & $-0.5013 \pm 0.0524$\\ \cline{3-5}
\hline
\multirow{3}{*}{$2^+$} & \multirow{3}{*}{$~~0.632$} & Without Noise   &$0.6309 \pm0.0212$  & $ 0.6321\pm 0.0002$   \\ \cline{3-5}
& & With Noise &  $0.6021 \pm 1.1771$  & $0.6211 \pm 0.7898$ \\ \cline{3-5}
\hline
\end{tabular}
\end{table*}

\begin{table}
\centering
\caption{Comparison of number of parameters P($\theta$) and gate count in the ansätze (Appendix~\ref{app:result_ansatz}) used for low lying states of $^{6}$Li in GC scheme.}
\label{tab:gate_li_gc}
\renewcommand{\arraystretch}{2}
\begin{tabular}{|c|c|c|c|c|c|c|}
\hline
\textbf{$J^{\pi}$} & \multicolumn{3}{c|}{\textbf{ADAPT-VQE}} & \multicolumn{3}{c|}{\textbf{VQE}} \\
\cline{2-7}
 & P($\theta$) & 1Q & 2Q & P($\theta$) & 1Q & 2Q \\
\hline
$0^+$ & 6 & 16 & 6 & 9 & 9 & 7 \\
\hline
$1^+$ & 4 & 14 & 4 & 7 & 7 & 4  \\
\hline
$2^+$ & 3 & 13 & 4 & 3 & 3 & 1  \\
\hline
$3^+$ & 1 & 2 & 0 & 1 & 1  & 0  \\
\hline
\end{tabular}
\end{table}

Calculations for $^6$Li are exclusively performed using  GC encoding, as JW requires 12 qubits and a significantly larger number of gates. This would significantly decrease the accuracy, as seen in the smaller case of $^{38}$Ar, and substantially increase the computational overhead. Therefore for nuclei with both valence protons and neutrons, GC encoding offers a far more practical and resource-efficient alternative. 

\begin{figure}
   \centering
    \includegraphics[width = \linewidth]{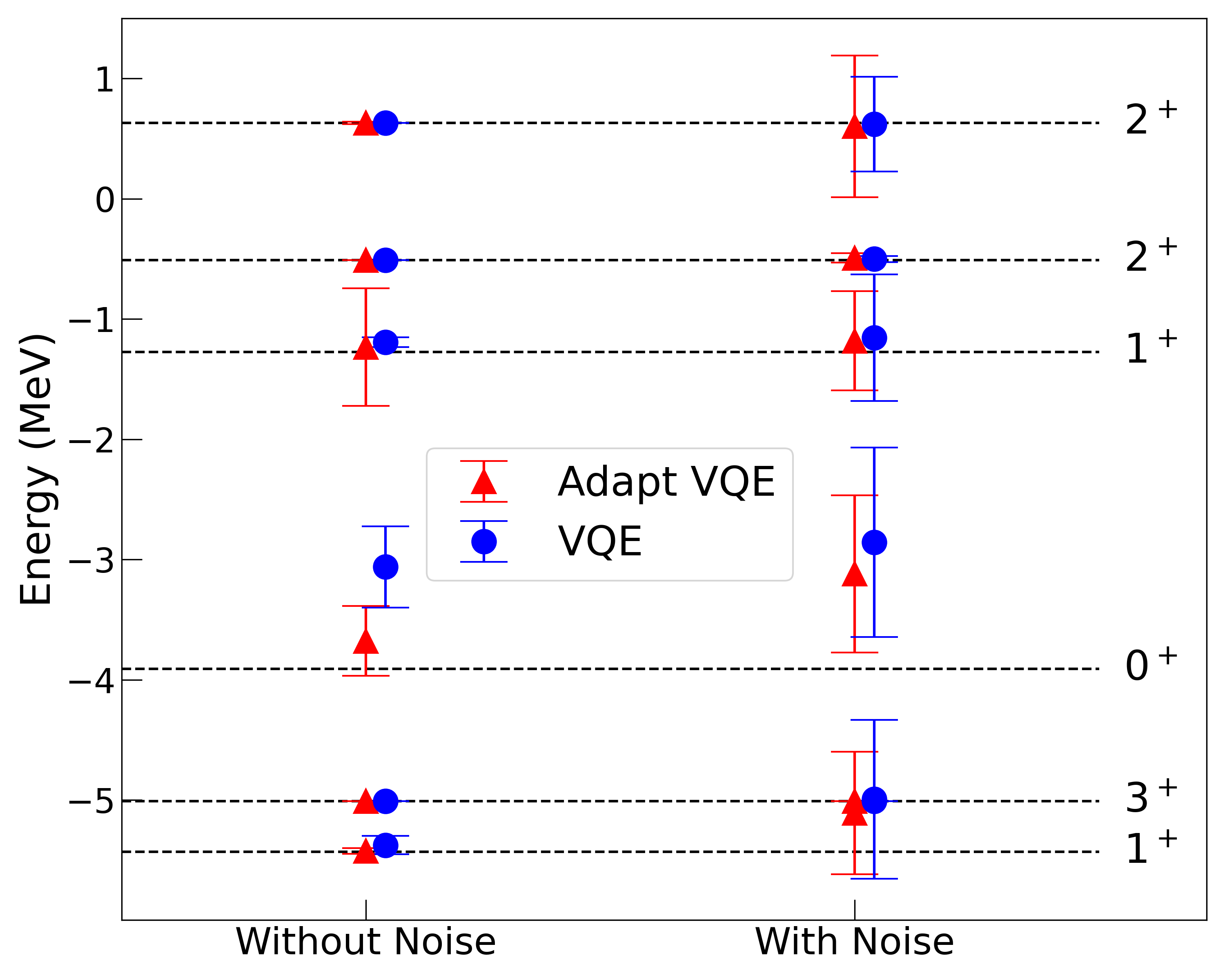}
    \caption{Level scheme of $^6$Li calculated using ADAPT-VQE and VQE, both with and without noise. The black dashed lines show the exact energy value for the state ($J^{\pi}$). }
    \label{fig:gc_li}
\end{figure}

The results for $^6$Li using GC encoding are given in Table~\ref{tab:Li_gc_table_0}. Qubit-ADAPT-VQE consistently outperforms VQE, particularly for $0^+$ and $1^+$ states, where the dynamically constructed ansätze has reduced circuit depth and width. For the $2^+$ and $3^+$ states, the results from both methods are nearly comparable, with VQE showing slightly better performance due to lesser gate count (Table~\ref{tab:gate_li_gc}). The complete level scheme is shown in Fig.~\ref{fig:gc_li}, comparing noiseless and noisy results for both algorithms. Under noiseless conditions, both methods reproduce the exact energy accurately, with ADAPT-VQE providing slightly better accuracy and smaller error bars for the $0^+$ state. However, when quantum noise is introduced, the calculated energies deviate significantly from the exact values. The uncertainty in measurement shown by the error bar also increases. Despite this, the ADAPT-VQE gives more accurate results for $0^+$ and $1^+$ states, highlighting its advantage for states with larger qubits. This pattern can be understood by seeing the ansätze complexity summarized in Table~\ref{tab:gate_li_gc}, which gives the number of parameters and gate counts required for both methods. For states like $0^+$ and $1^+$, ADAPT-VQE manages to achieve better performance with much fewer parameters and fewer two-qubit gates ($2$Q) than VQE. While the number of single-qubit gates ($1$Q) is slightly higher in the ADAPT-VQE case, but these have a relatively minor impact on fidelity compared to $2$Q gates and parameters. For single qubit states like $3^+$, where both ansätze are extremely shallow and involve only a few gates, the performance of VQE and ADAPT-VQE are identical. 

Compared to $^{38}$Ar, the accuracy for $^6$Li is slightly lower due to the larger number of parameters and gates. However, the use of GC encoding significantly reduced the resource requirement and computational time compared to JW, making it the best option for nuclei with a large number of valance nucleons. The same optimizer (SPSA), number of shots, and number of runs are used as in the $^{38}$Ar calculations to ensure a fair and direct comparison.



\subsection{Zero-noise extrapolation}
\begin{figure}[h!]
   \centering
    \includegraphics[width = \linewidth]{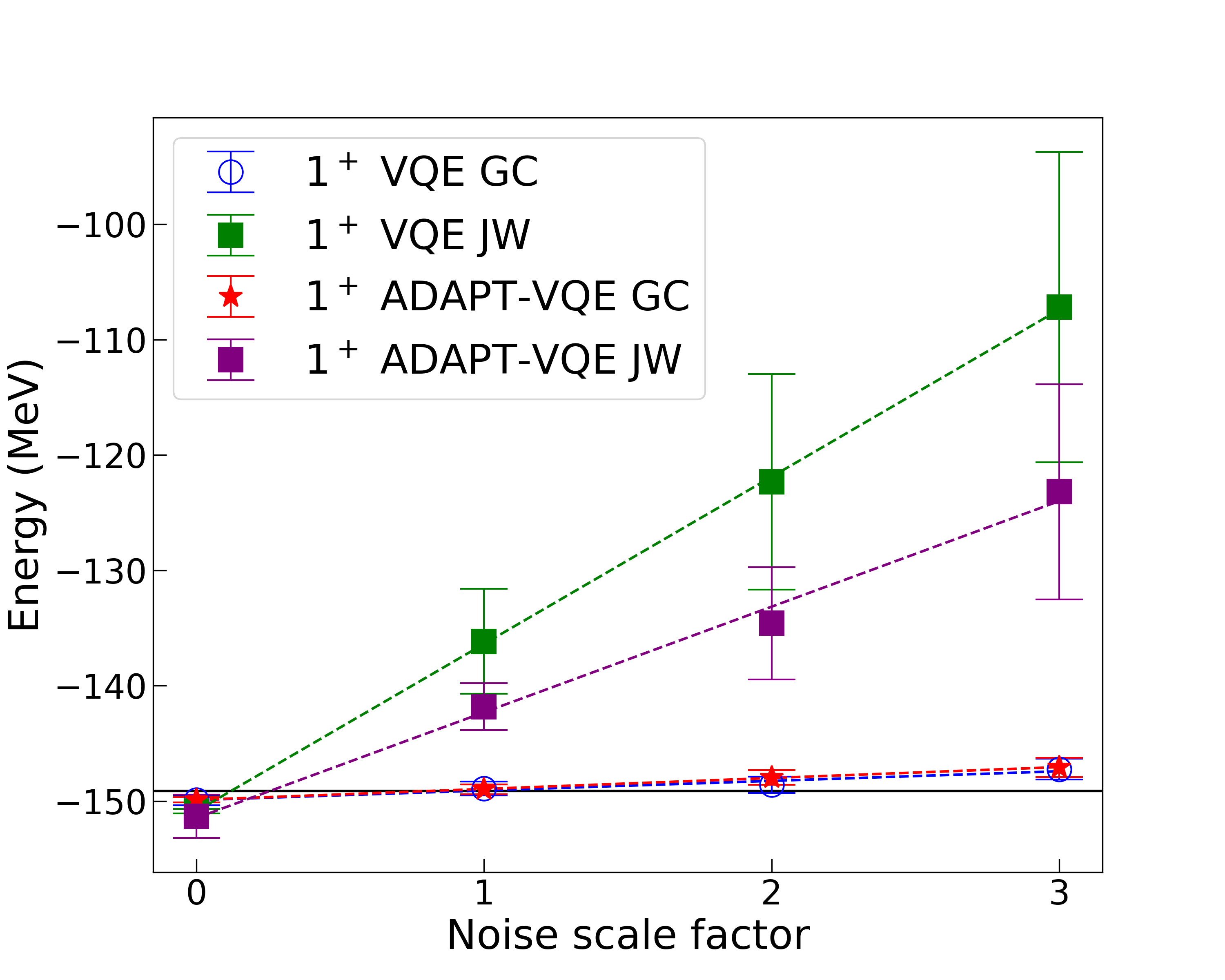}
    \caption{Zero noise extrapolation for $1^+$ state of $^{38}$Ar. The raw with noise results are shown at noise scale factor one, while two-fold and three-fold noise scaled results are extrapolated back to the zero noise limit. The black solid line shows the exact value of $-149.104$ MeV. Results are shown for VQE and ADAPT-VQE using both JW and GC schemes. }
    \label{fig:zne_1}
\end{figure}

\begin{figure}[h!]
   \centering
    \includegraphics[width = \linewidth]{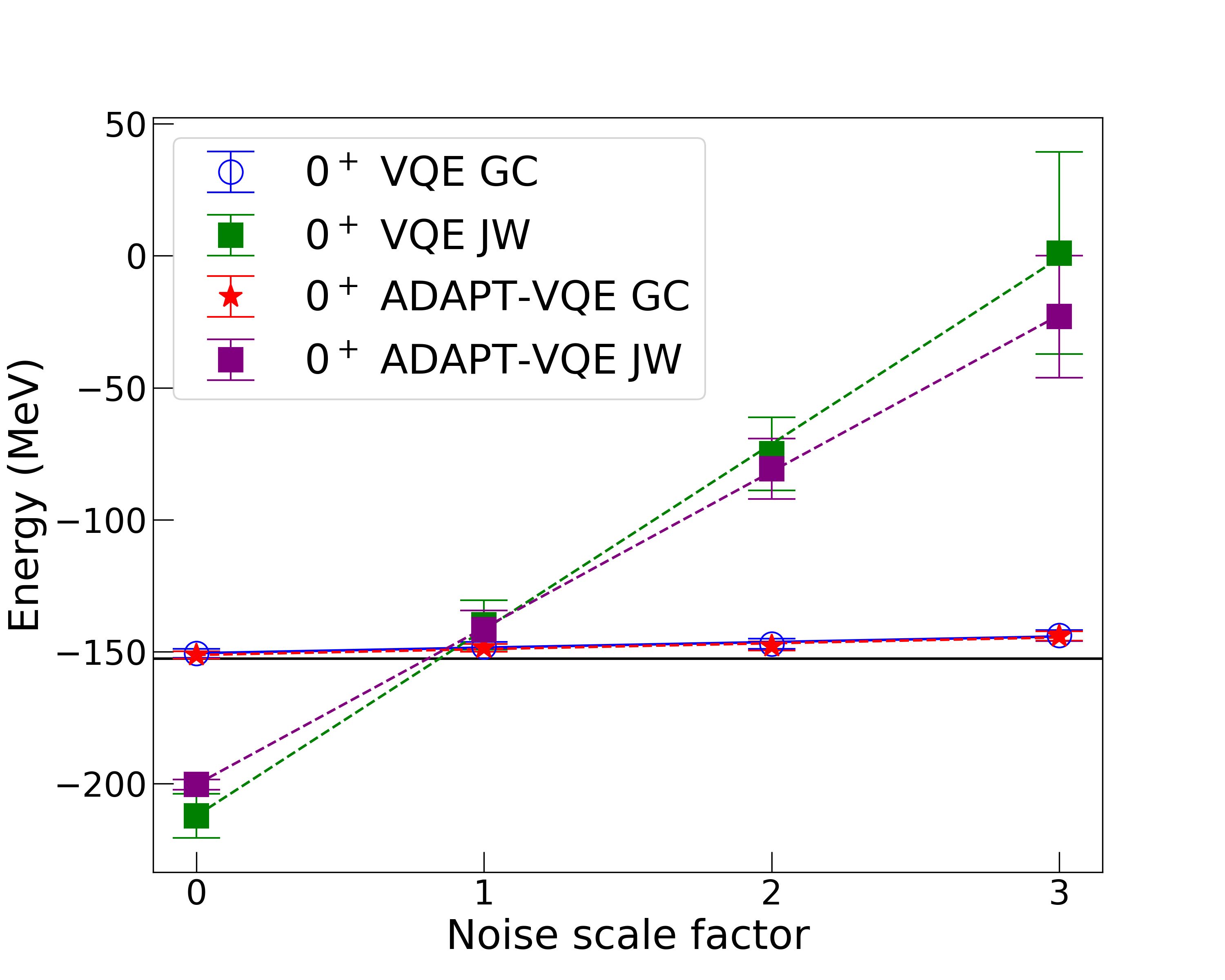}
    \caption{Same as Fig.~\ref{fig:zne_1} but for $0^+$ state.}
    \label{fig:zne_0}
\end{figure}

To mitigate the errors due to imperfections in hardware and gate operations, we employ the zero-noise extrapolation (ZNE)~\cite{zne1} technique using the unitary folding method. It mitigates the noise by intentionally increasing the noise level and then extrapolating the results to the zero noise limit. In the unitary folding method, we increase the depth of the circuit through gate folding where each gate $G$ is replaced with $G G^{\dagger} G$. Using this, we have computed the ground state energies of $1^+$ and $0^+$ state of $^{38}$Ar. These particular states were chosen because their ansätze involves a deeper circuit. This makes them suitable for studying the impact of noise and the effectiveness of ZNE. In contrast, the $2^+$ state of $^{38}$Ar is represented by a single qubit ansatz where the gate folding does not significantly affect the noise, and hence was not included for comparison. The linearly extrapolated results using different encoding schemes and algorithms are given in Tables ~\ref{tab:zne1+} and  ~\ref{tab:zne0+}. Additionally, the energy values for different noise scale factors and linear fit are shown in Fig.~\ref{fig:zne_1} and 
~\ref{fig:zne_0}.

From Table ~\ref{tab:zne1+}, it is evident that the GC encoding yields the extrapolated energies closer to the exact value. However, the JW shows a greater deviation, suggesting it is more sensitive to noise due to increased circuit depth and gate errors. For $0^+$ state, the JW scheme results in a significantly large deviation. Overall the results show that GC has better accuracy, where extrapolated values are much closer to the exact values.  

\begin{table}
\centering
\caption{Linearly extrapolated results in zero-noise limit for $1^+$ of $^{38}$Ar. The exact energy value is $-149.104$.}
\label{tab:zne1+}
\setlength{\tabcolsep}{8pt}
\renewcommand{\arraystretch}{1.6}
\begin{tabular}{|c|c|l|c|c|}
\hline
\textbf{Encodings} & \textbf{ADAPT-VQE } & \textbf{VQE} \\ \hline
GC &$  -149.8854 \pm 0.2280$ &$ -149.9072 \pm 0.4460$  \\ \hline
JW &$-151.3025 \pm 1.8803 $  &$  -150.8633 \pm 0.2063$ \\ \hline
\end{tabular}
\end{table}

\begin{table}
\centering
\caption{Linearly extrapolated results in zero-noise limit for $0^+$ of $^{38}$Ar. The exact energy value is $-152.677$.}
\label{tab:zne0+}
\setlength{\tabcolsep}{8pt}
\renewcommand{\arraystretch}{1.6}
\begin{tabular}{|c|c|l|c|c|}
\hline
\textbf{Encodings} & \textbf{ADAPT-VQE } & \textbf{VQE} \\ \hline
GC &$-151.2478 \pm1.3394 $ &$ -150.5822 \pm 1.6827$  \\ \hline
JW &$ 200.3508 \pm 1.9614$  &$  -212.1988 \pm 8.3705$ \\ \hline
\end{tabular}
\end{table}

\section{Conclusions}\label{sec:Conclusions}
We have presented the noise-resilient protocols to perform the nuclear shell model calculations on a quantum computer by significantly reducing the resource requirements in terms of qubits and gates. We have provided for the first time a scheme to map the many-body states on qubits using the GC encoding and transform the fermionic operators, preserving the antisymmetrization of the wave function under the exchange of particles. Comparing with the conventional JW scheme to map the fermionic states on qubits, GC performs much better under all circumstances, that is, different algorithms like VQE and qubit-ADAPT-VQE, noise, and different nuclei. This not only enhances the accuracy of results but also makes the calculations more practical for simulations of complex nuclear systems, especially for nuclei for which both protons and neutrons act as valance particles. Since this encoding incorporates all the available states with a lesser number of qubits, it allows us to use hardware-efficient variational ansatz for VQE. This eliminates the need to construct particle-conserving ansatz as required in JW. Additionally, the dynamically created qubit-ADAPT-VQE ansatz using the GC scheme is also much smaller as compared to those generated for JW. This reduction in qubits and gates is essential for efficient quantum computing, particularly for large-scale simulations.

Furthermore, we have explored the performance and resource requirement of two quantum algorithms, VQE and qubit-ADAPT-VQE, using both encoding schemes, GC and JW, and various ansätze designs. We applied these algorithms to calculate the energy spectra of nuclei $^{38}$Ar and $^6$Li using the nuclear shell model. Our results show the advantage of using qubit-ADAPT-VQE dynamically constructed ansätze, which provides significant improvement in accuracy and resource requirement compared to standard VQE, especially for cases involving large circuit depth and width. 

One more major contribution of this work is the optimization of particle conserving ansatz. We have given a method to construct the particle-conserving ansatz with fewer gates and reduced depth compared to existing approaches. This improvement not only reduces the computational overhead but also enhances scalability as the difference in gate count becomes more noticeable with an increasing number of qubits. 

Overall, this work highlights the potential of resource-efficient encodings and hybrid quantum-classical approaches to scale the quantum simulations to larger systems. This advantage brings us closer to practical quantum simulations of nuclear systems and allows us to handle more complex nuclei and interactions. 

\begin{acknowledgments}
The Science and Engineering Research Board partly supported this work under Grant Code  CRG/2022/009359. This work was performed under the auspices of the U.S. Department of Energy by Lawrence Livermore National Laboratory under Contract No. DE-AC52-07NA27344. We acknowledge the National Supercomputing Mission (NSM) for providing computing resources of ‘PARAM Ganga’ at the Indian Institute of Technology Roorkee, which is implemented by C-DAC and supported by the Ministry of Electronics and Information Technology (MeitY) and Department of Science and Technology (DST), Government of India. We thank Vidisha Aggarwal for
valuable discussions at the initial stage of this work, and Calvin Johnson for providing us the matrix elements for shell-model calculations.

\end{acknowledgments}

\begin{appendix}
\section{Transformation under Gray code encoding}\label{app:encoding}
Consider a system of three particles distributed across five orbitals. The total number of particle-conserving states in this configuration is 10. However, assume that only 8 of these states contribute to the Hamiltonian. Using Gray code encoding, these 8 states can be efficiently mapped onto just 3 qubits. The corresponding mapping of states is shown in Table~\ref{tab:encoding}. To transform the one- and two-body operators comprising the Hamiltonian in terms of Pauli matrices, we use the expressions given in section~\ref{sec:encoding}.

\begin{table}[h!]
    \centering
    \caption{Mapping of fermionic states in occupation number representation in one-hot (which is same as occupation number representation) and GC encoding.}
    \setlength{\tabcolsep}{8pt}
    \renewcommand{\arraystretch}{1.5}
    \begin{tabular}{|c|c|c|}
    \hline
       $\ket{\psi}$ & One-hot & GC \\
       \hline
       $\ket{0}$& $\ket{00111}$  & $\ket{000}$ \\
       $\ket{1}$& $\ket{01011}$  & $\ket{001}$ \\
       $\ket{2}$& $\ket{01101}$  & $\ket{011}$ \\
       $\ket{3}$& $\ket{01110}$  & $\ket{010}$ \\
       $\ket{4}$& $\ket{10011}$  & $\ket{110}$ \\
       $\ket{5}$& $\ket{10101}$  & $\ket{100}$ \\
       $\ket{6}$& $\ket{10110}$  & $\ket{101}$ \\
       $\ket{7}$& $\ket{11001}$  & $\ket{111}$ \\
    \hline
    \end{tabular}
    \label{tab:encoding}
\end{table}

For example, the operator $\ket{2}\bra{3}$ in different bases is given by
\begin{equation}
    \ket{2}\bra{3} = \ket{01101}\bra{01110}=c_1^{\dagger}c_0=\ket{011}\bra{010}
\end{equation}
which can be further transformed in terms of Paulis as
\begin{equation}
    \ket{011}\bra{010}=P^{(0)}\otimes P^{(1)}\otimes\sigma^{-}.
\end{equation}
Similarly
\begin{eqnarray}
    \ket{1}\bra{7} &=& \ket{01011}\bra{11001}=-c_4^{\dagger}c_1=-\ket{001}\bra{111}\nonumber\\
    &=& -\sigma^{+}\otimes\sigma^{+}\otimes P^{(1)}.
\end{eqnarray}
A two-body operator can be transformed as 
\begin{eqnarray}
    \ket{2}\bra{4} &=&\ket{01101}\bra{10011}=c_4^{\dagger}c_1^{\dagger}c_3c_2=\ket{011}\bra{110}\nonumber\\
    &=&\sigma^{+}\otimes P^{(1)}\otimes \sigma^{-}.
\end{eqnarray}
\section{Ansatz design}\label{app:ansatz}
Here, we present an illustration on preparing a particle-conserving ansatz using the technique given in Section~\ref{sec:ansatz}. Let us consider a system of 6 qubits and 2 particles and 4 basis states, $\ket{110000},\ket{001100},\ket{000011},$ and $\ket{001001}$. Our particle number conserving ansatz should have all and only these states as follows.
\begin{equation}\label{eq:ansatz}
    \ket{\psi}=c_1\ket{000011}+c_2\ket{001100}+c_3\ket{110000}+c_4\ket{100100}
\end{equation}

To prepare this ansatz, we initialize our circuit with the first basis state $\ket{000011}$ applying $X$ gate on first and second qubits. Then we create a superposition of first and second basis states $c_1\ket{000011}+c_2\ket{001100}$ by applying a double excitation circuit on 1st, 2nd, 3rd and 4th qubits. Next, to create a superposition of 1st, 2nd, and 3rd basis states, we consider 2nd and 3rd states only and apply a double excitation gate on 3rd, 4th, 5th, and 6th states only. But since we should not change the second state itself, we apply control gates on the 3rd and 4th qubits as these are one in the second state. Similarly, going from the 3rd to 4th states, a single-excitation circuit is required to be applied on the 3rd and 5th qubits with control on the 6th qubit as this should remain 1.

\begin{figure}[h!]
\begin{tikzpicture}
\node[scale=0.9] {
\begin{quantikz}
\ket{0}&\gate{X}&\gate[wires=4][0.5cm]{G_{2}^{(1234)}(\theta_1)}&\qw&\qw&\qw&\qw\\
\ket{0}&\gate{X}&\qw&\qw&\qw&\qw&\qw\\
\ket{0}&\qw&\qw&\gate[wires=4][0.5cm]{G_{2}^{(3456)}(\theta_2)}&\qw&\gate[wires=4][0.5cm]{G_{1}^{(356)}(\theta_3)}&\qw\\
\ket{0}&\qw&\qw&\qw&\qw&\qw&\qw\\
\ket{0}&\qw&\qw&\qw&\qw&\qw&\qw\\
\ket{0}&\qw&\qw&\qw&\qw&\qw&\qw
\end{quantikz}
};
\end{tikzpicture}

\caption{Quantum circuit to prepare ansatz given in Eq.~\eqref{eq:ansatz} with $G_1$ and $G_2$ as the single- and double excitation operators expanded in subsequent figures.}\label{fig:ansatz}
\end{figure}
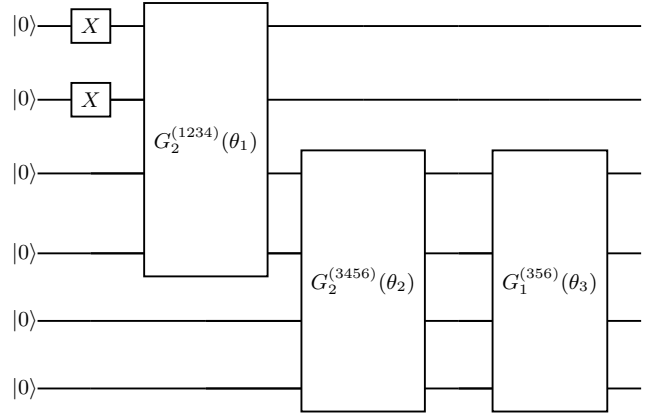

where

\begin{figure}[h!]
\begin{tikzpicture}
\node[scale=0.9] {
\begin{quantikz}
\qw&\gate[wires=4][0.5cm]{G_{2}^{(1234)}(\theta_1)}&\qw\\
\qw&\qw&\qw\\
\qw&\qw&\qw\\
\qw&\qw&\qw
\end{quantikz}
$\equiv$
\begin{quantikz}
\qw&\qw&\qw&\qw&\targ{}&\qw\\
\qw&\qw&\qw&\targ{}&\qw&\qw\\
\qw&\qw&\targ{}&\qw&\qw&\qw\\
\qw&\gate{Y(\theta_1)}&\ctrl{-1}&\ctrl{-2}&\ctrl{-3}&\qw
\end{quantikz}
};
\end{tikzpicture}

\end{figure}

\begin{figure}[h!]
\begin{tikzpicture}
\node[scale=0.9] {
\begin{quantikz}
\qw&\gate[wires=4][0.5cm]{G_{2}^{(3456)}(\theta_2)}&\qw\\
\qw&\qw&\qw\\
\qw&\qw&\qw\\
\qw&\qw&\qw
\end{quantikz}
$\equiv$
\begin{quantikz}
\qw&\ctrl{3}&\qw&\qw&\targ{}&\qw\\
\qw&\ctrl{2}&\qw&\targ{}&\qw&\qw\\
\qw&\qw&\targ{}&\qw&\qw&\qw\\
\qw&\gate{Y(\theta_2)}&\ctrl{-1}&\ctrl{-2}&\ctrl{-3}&\qw
\end{quantikz}
};
\end{tikzpicture}

\end{figure}

\begin{figure}[h!]
\begin{tikzpicture}
\node[scale=0.9] {
\begin{quantikz}
\qw&\gate[wires=4][0.5cm]{G_{1}^{(356)}(\theta_2)}&\qw\\
\qw&\qw&\qw\\
\qw&\qw&\qw\\
\qw&\qw&\qw
\end{quantikz}
$\equiv$
\begin{quantikz}
\qw&\gate{Y(\theta_3)}&\ctrl{2}&\qw\\
\qw&\qw&\qw&\qw\\
\qw&\ctrl{-2}&\targ{}&\qw\\
\qw&\ctrl{-1}&\ctrl{-1}&\qw
\end{quantikz}
};
\end{tikzpicture}

\end{figure}
These circuits can further be decomposed in terms of single- and two-qubit gates.

\section{$^6$Li}
In this section, we show how the basis states for different low-lying spin-parity states of $^6$Li are mapped onto the qubit states in both JW and GC encoding schemes, as given in Tables~\ref{tab:my_label_0+},~\ref{tab:my_label_1+},~\ref{tab:my_label_2+}, and~\ref{tab:my_label_3+}.

\begin{table}[h!]
    \centering
    \caption{Mapping of $0^+$.}
    \label{tab:my_label_0+}
    \renewcommand{\arraystretch}{1.6}
    \begin{tabular}{|r|c|c|}
    \hline
    $\ket{\psi}_f$ & JW & GC \\
    \hline
    $\ket{1}$  &  $\ket{100000000100}$ &  $\ket{0000}$ \\
    $\ket{2}$  &  $\ket{010000000010}$ &  $\ket{0001}$ \\
    $\ket{3}$  &  $\ket{001000000010}$ &  $\ket{0011}$ \\
    $\ket{4}$  &  $\ket{010000000001}$ &  $\ket{0010}$ \\
    $\ket{5}$  &  $\ket{001000000001}$ &  $\ket{0110}$ \\
    $\ket{6}$  &  $\ket{000010010000}$ &  $\ket{0100}$ \\
    $\ket{7}$  &  $\ket{000001010000}$ &  $\ket{1100}$ \\
    $\ket{8}$  &  $\ket{000010001000}$ &  $\ket{1000}$ \\
    $\ket{9}$  &  $\ket{000001001000}$ &  $\ket{1001}$ \\
    $\ket{10}$ &  $\ket{000100100000}$ &  $\ket{1101}$ \\
\hline
    \end{tabular}
\end{table}

\begin{table}[h!]
    \centering
    \caption{Mapping of $1^+$.}
    \label{tab:my_label_1+}
    \renewcommand{\arraystretch}{1.6}
    \begin{tabular}{|r|c|c|}
    \hline
    $\ket{\psi}_f$ & JW & GC\\
    \hline
    $\ket{1}$  &  $\ket{100000000010}$ &  $\ket{000}$ \\
    $\ket{2}$  &  $\ket{100000000001}$ &  $\ket{001}$ \\
    $\ket{3}$  &  $\ket{010000010000}$ &  $\ket{011}$ \\
    $\ket{4}$  &  $\ket{001000010000}$ &  $\ket{010}$ \\
    $\ket{5}$  &  $\ket{010000001000}$ &  $\ket{110}$ \\
    $\ket{6}$  &  $\ket{001000001000}$ &  $\ket{100}$ \\
    $\ket{7}$  &  $\ket{000010100000}$ &  $\ket{101}$ \\
    $\ket{8}$  &  $\ket{000001100000}$ &  $\ket{111}$ \\
\hline
    \end{tabular}
\end{table}

\begin{table}[h!]
    \centering
    \caption{Mapping of $2^+$.}
    \label{tab:my_label_2+}
    \renewcommand{\arraystretch}{1.6}
    \begin{tabular}{|r|c|c|}
    \hline
    $\ket{\psi}_f$ & JW & GC\\
    \hline
    $\ket{1}$  &  $\ket{100000010000}$ &  $\ket{00}$ \\
    $\ket{2}$  &  $\ket{100000001000}$ &  $\ket{01}$ \\
    $\ket{3}$  &  $\ket{010000100000}$ &  $\ket{11}$ \\
    $\ket{4}$  &  $\ket{001000100000}$ &  $\ket{10}$ \\
    \hline
    \end{tabular}
\end{table}

\begin{table}[h!]
    \centering
    \caption{Mapping of $3^+$.}
    \label{tab:my_label_3+}
    \renewcommand{\arraystretch}{1.6}
    \begin{tabular}{|r|c|c|}
    \hline
    $\ket{\psi}_f$ & JW & GC\\
    \hline
    $\ket{1}$  &  $\ket{100000100000}$ &  $\ket{0}$ \\
    \hline
    \end{tabular}
\end{table}

\section{$^{38}$Ar}
Here we present the mapping of basis states for various spin-parity configurations of $^{38}$Ar using both JW and GC encodings, as given in Tables~\ref{tab:my_label_0},~\ref{tab:my_label_1}, and~\ref{tab:my_label_2}.


\begin{table}[h!]
    \centering
    \caption{Mapping of $0^+$.}
    \label{tab:my_label_0}
    \renewcommand{\arraystretch}{1.6}
    \begin{tabular}{|r|c|c|}
    \hline
    $\ket{\psi}_f$ & JW & GC \\
    \hline
    $\ket{1}$  &  $\ket{010010}$ &  $\ket{000}$ \\
    $\ket{2}$  &  $\ket{001010}$ &  $\ket{001}$ \\
    $\ket{3}$  &  $\ket{010001}$ &  $\ket{011}$ \\
    $\ket{4}$  &  $\ket{001001}$ &  $\ket{010}$ \\
    $\ket{5}$  &  $\ket{100100}$ & $\ket{110}$ \\
\hline
    \end{tabular}
\end{table}

\begin{table}[h!]
    \centering
    \caption{Mapping of $1^+$.}
    \label{tab:my_label_1}
    \renewcommand{\arraystretch}{1.6}
    \begin{tabular}{|r|c|c|}
    \hline
    $\ket{\psi}_f$ & JW & GC \\
    \hline
    $\ket{1}$  &  $\ket{100010}$ &  $\ket{00}$ \\
    $\ket{2}$  &  $\ket{100001}$ &  $\ket{01}$ \\
    $\ket{3}$  &  $\ket{011000}$ &  $\ket{11}$ \\
\hline
    \end{tabular}
\end{table}

\begin{table}[h!]
    \centering
    \caption{Mapping of $2^+$.}
    \label{tab:my_label_2}
    \renewcommand{\arraystretch}{1.6}
    \begin{tabular}{|r|c|c|}
    \hline
    $\ket{\psi}_f$ & JW & GC \\
    \hline
    $\ket{1}$  &  $\ket{110000}$ &  $\ket{0}$ \\
    $\ket{2}$  &  $\ket{101000}$ &  $\ket{1}$ \\
\hline
    \end{tabular}
\end{table}

\section{Transformed Hamiltonian}\label{app:transformed_ham}
In this appendix, we provide the explicit form of the transformed Hamiltonian used in our quantum simulations. For each low-lying spin-parity states of $^{38}$Ar and $^{6}$Li, we have listed coefficients along with their corresponding tensor product operators in both JW and GC schemes, as given in Tables~\ref{tab:H0+ligc}, ~\ref{tab:H1+ligc},  ~\ref{tab:H2+ligc},  ~\ref{tab:H3+ligc},  ~\ref{tab:H0+argc},  ~\ref{tab:H1+argc},  ~\ref{tab:H2+argc},  ~\ref{tab:H0+arjw},  ~\ref{tab:H1+arjw}, and~\ref{tab:H2+arjw}.

\begin{table*}[h]
    \centering
    \caption{The coefficients and tensor product operators of the Hamiltonian $H_{0^+}$ in GC scheme for $^{6}$Li.}
    \label{tab:H0+ligc}
    \renewcommand{\arraystretch}{1.2}
    \begin{tabular}{|c|c|c|c|c|c|}
        \hline
        \textbf{Operator} & \textbf{Coefficient} & \textbf{Operator} & \textbf{Coefficient}& \textbf{Operator} & \textbf{Coefficient} \\
        \hline
         $ I $&$\phantom{-}0.242340$ & $  X_3 $&$\phantom{-}0.026711$&$ Y_2Y_3Z_0Z_1  $& $- 0.506821$ \\
         $  Z_3  $&$- 0.051668$ & $  X_2 $&$- 0.135605$ & $ Y_1Y_2Z_0Z_3  $&$- 0.210596$\\
         $ Z_2  $& $- 0.351740$&$  X_2X_3 $ & $\phantom{-}0.149642$& $ Y_1Y_3Z_0Z_2 $&$\phantom{-}0.318451$\\ 
         $  X_2Z_3  $& $- 0.135605$&$  X_3Z_2 $ & $- 0.026711$&$ Y_0Y_2Z_1Z_3  $& $- 0.135605$\\
         $  Y_2Y_3  $&$- 0.149642$ &$  Z_2Z_3 $ & $- 0.057731$& $ X_3Z_0Z_1Z_2 $&$\phantom{-}0.133192$\\
         $  X_1  $&$\phantom{-}0.026711$ &$ Z_1 $ &$- 0.057731$&$ Y_0Y_3Z_1Z_2 $ & $- 0.133192$ \\
         $  X_1X_3  $& $- 0.248567$&$  X_1Z_3 $ & $\phantom{-}0.026711$& $ Y_0Y_1Z_2Z_3 $&$\phantom{-}0.026711$\\
         $  X_3Z_1  $&$\phantom{-}0.026711$ &$  Y_1Y_3 $ &$- 0.248567$& $ Y_0Y_1Y_2Y_3  $&$- 0.242086$  \\
         $  Z_1Z_3  $& $- 0.351740$&$  X_1X_2 $ &$- 0.408183$& $ X_0Z_1Z_2Z_3  $ &$- 0.615660$ \\
         $  X_1Z_2  $&$\phantom{-}0.331685$ &$  X_2Z_1 $ & $- 0.222791$  & $ X_1X_2Z_0Z_3 $&$0.210596$ \\
         $  Y_1Y_2  $& $\phantom{-}0.486446$&$  Z_1Z_2 $ &$- 0.446855$ & $ X_1X_2X_3Z_0 $& $\phantom{-}0.242086$ \\
         $  X_1X_2X_3  $&$- 0.135605$ &$  X_1X_2Z_3 $ & $- 0.408183$ & $ X_1X_2Y_0Y_3  $&$- 0.242086$\\
         $  X_1X_3Z_2  $ & $\phantom{-}0.248567$&$  X_1Y_2Y_3 $ &$\phantom{-}0.135605$& $ X_1X_3Y_0Y_2  $&$- 0.242086$ \\
         $  X_1Z_2Z_3  $&$\phantom{-}0.331685$ & $  X_2X_3Z_1 $&$\phantom{-}0.149642$  & $ X_1X_3Z_0Z_2 $&$\phantom{-}0.318451$\\
         $ X_2Y_1Y_3  $&$- 0.135605$ &$  X_2Z_1Z_3 $ & $- 0.222791$& $ X_1Y_0Y_2Z_3  $&$\phantom{-}0.366906$\\
         $  X_3Y_1Y_2  $&$- 0.135605$ & $  X_3Z_1Z_2 $&$- 0.026711$ &$ X_1Y_0Y_3Z_2 $ &$\phantom{-}0.139914$ \\
         $  Y_1Y_2Z_3  $&$\phantom{-}0.486446$ & $  Y_1Y_3Z_2 $& $\phantom{-}0.248567$ & $  X_1Y_2Y_3Z_0  $& $- 0.242086$\\
         $  Y_2Y_3Z_1  $& $- 0.149642$&$  Z_1Z_2Z_3 $ &$- 0.152846$ & $  X_1Z_0Z_2Z_3 $&$- 0.065300$ \\
         $ X_0  $&$- 0.067722$ &$  Z_0 $ & $\phantom{-}0.601002$ & $ X_2X_3Y_0Y_1  $&$\phantom{-}0.242086$\\
         $  X_0X_3 $&$- 0.026711$ & $  X_0Z_3 $&$\phantom{-}0.168995$ & $ X_2Y_0Y_1Z_3  $&$\phantom{-}0.366906$  \\
         $  X_3Z_0  $& $- 0.133192$&$  Y_0Y_3 $ &$\phantom{-}0.133192$&$ X_2Y_0Y_3Z_1 $ & $- 0.120848$ \\
         $  Z_0Z_3  $&$\phantom{-}0.191529$ &$  X_0X_2 $ & $\phantom{-}0.135605$& $ X_2X_3Z_0Z_1 $&$\phantom{-}0.506821$ \\
         $  X_0Z_2  $& $\phantom{-}0.411575$&$  X_2Z_0 $ & $- 0.135605$ & $ X_2Y_1Y_3Z_0  $& $\phantom{-}0.242086$\\
         $  Y_0Y_2  $&$- 0.261380$ &$  Z_0Z_2 $ &$- 0.409472$ & $ X_2Z_0Z_1Z_3 $& $- 0.261380$ \\
         $ X_0X_2X_3  $&$- 0.395699$ & $  X_0X_2Z_3 $&$\phantom{-}0.135605$ & $ X_3Y_0Y_1Z_2  $&$- 0.139914$\\
         $  X_0X_3Z_2  $&$\phantom{-}0.026711$ & $ X_0Y_2Y_3 $&$\phantom{-}0.395699$& $ X_3Y_0Y_2Z_1 $&$- 0.120848$ \\  $ X_0Z_2Z_3  $&$\phantom{-}0.174856$ &$ X_2X_3Z_0 $ &$\phantom{-}0.506821$ & $ X_0X_1X_3Z_2 $&$- 0.516548$ \\
         $ X_2Y_0Y_3  $&$- 0.120848$ &$ X_2Z_0Z_3 $ & $- 0.135605$&$ X_0X_1Y_2Y_3  $ & $- 0.135605$\\
         $ X_3Y_0Y_2  $&$- 0.120848$ &$ X_3Z_0Z_2 $ & $\phantom{-}0.133192$ & $ X_0X_1Z_2Z_3 $&$- 0.331685$ \\
         $ Y_0Y_2Z_3  $&$- 0.261380$ & $ Y_0Y_3Z_2 $&$- 0.133192$  &$ X_0X_2X_3Z_1  $ & $- 0.395699$\\
         $ Y_2Y_3Z_0 $&$- 0.506821$ & $ X_0X_1 $&$- 0.026711$ & $ X_0X_2Y_1Y_3 $&$\phantom{-}0.135605$\\
         $ X_0Z_1  $&$\phantom{-}0.174856$ & $ X_1Z_0 $&$\phantom{-}0.026711$ &$ X_0X_2Z_1Z_3  $ & $\phantom{-}0.222791$\\
         $ Y_0Y_1  $ &$- 0.065300$ & $ X_3Y_1Y_2Z_0  $& $\phantom{-}0.242086$& $ X_0X_3Y_1Y_2 $& $\phantom{-}0.135605$ \\
         $ X_0X_1X_3 $&$\phantom{-}0.516548$ &$ X_0X_1Z_3  $ &$- 0.026711$& $ X_0X_3Z_1Z_2  $&$\phantom{-}0.026711$  \\
         $  X_0X_3Z_1 $ &$- 0.026711$ & $  X_0Y_1Y_3  $&$\phantom{-}0.516548$ & $ X_0Y_1Y_2Z_3 $&$- 0.647896$  \\
         $ X_0Z_1Z_3 $&$\phantom{-}0.411575$ &$ X_1X_3Z_0  $ &$- 0.318451$ & $ X_0Y_1Y_3Z_2  $& $- 0.516548$ \\
         $ X_1Y_0Y_3 $ &$- 0.139914$ & $ X_1Z_0Z_3  $&$\phantom{-}0.026711$  & $ X_0Y_2Y_3Z_1 $ &$\phantom{-}0.395699$\\
         $ X_3Y_0Y_1 $&$\phantom{-}0.139914$ & $ X_3Z_0Z_1  $&$- 0.133192$ & $ Z_0Z_1Z_2Z_3 $&$\phantom{-}0.601002$ \\
         $ Y_0Y_1Z_3 $&$- 0.065300$ &$ Y_0Y_3Z_1  $ &$\phantom{-}0.133192$ & $ X_0X_1X_2X_3 $&$\phantom{-}0.135605$ \\
         $ Y_1Y_3Z_0 $&$- 0.318451$ &$ Z_0Z_1Z_3  $ &$- 0.409472$ &$ X_0X_1X_2Z_3  $ &$- 0.628441$ \\
         $ X_0X_1X_2 $&$- 0.628441$ &$ X_0X_1Z_2  $ &$- 0.331685$& $ Y_0Y_1Z_2 $&$\phantom{-} 0.026711$   \\
         $ X_0X_2Z_1 $&$\phantom{-}0.222791$ & $ X_0Y_1Y_2  $&$-0.647896$& $ Z_0Z_1Z_2 $&$\phantom{-}0.191529$  \\
         $ X_0Z_1Z_2 $& $- 0.378941$&$ X_1X_2Z_0  $ & $\phantom{-}0.210596$ & $ Y_1Y_2Z_0 $&$- 0.210596$\\
         $ X_1Y_0Y_2 $& $\phantom{-}0.366906$&$ X_1Z_0Z_2 $ & $- 0.065300$&$ Y_0Y_2Z_1  $ &$- 0.135605$ \\
         $ X_2Y_0Y_1 $& $\phantom{-}0.366906$&$ X_2Z_0Z_1  $ &$-0.261380$& &\\
         \hline

\end{tabular}
\end{table*}

\begin{table*}[h]
    \centering
    \caption{The coefficients and tensor product operators of the Hamiltonian $H_{1^+}$ in GC scheme for $^{6}$Li.}
    \label{tab:H1+ligc}
    \renewcommand{\arraystretch}{1.2}
    \begin{tabular}{|c|c|c|c|c|c|}
        \hline
        \textbf{Operator} & \textbf{Coefficient} & \textbf{Operator} & \textbf{Coefficient} & \textbf{Operator} & \textbf{Coefficient} \\
        \hline
         $ I$&$-0.016731$ &$  X_2$ & $-0.025631$&$ Z_0Z_1Z_2$ &$-0.726409$ \\
         $  Z_2 $&$-0.502801$ &$ X_1$ & $\phantom{-}0.202641$ & $ Y_1Y_2Z_0$ &$\phantom{-}0.893859$ \\
         $ Z_1 $&$\phantom{-}0.281627$ & $ X_1X_2$&$-0.457219$ & $ Y_0Y_2Z_1$&$-0.428803$ \\
         $ X_1Z_2 $&$-0.756565$ &$ X_2Z_1$ & $ \phantom{-}1.536397$ & $ Y_0Y_1Z_2 $&$\phantom{-}0.720015$ \\
         $ Y_1Y_2 $&$-0.138482$ &$ Z_1Z_2$ &$ \phantom{-}0.502801$ &$ X_2Z_0Z_1$ & $\phantom{-}0.221354$\\
         $ X_0 $& $\phantom{-}0.392524$&$ Z_0$ &$ 0.583498$ & $ X_2Y_0Y_1  $& $-0.241870$\\
         $ X_0X_2 $&$-1.334465$ & $ X_0Z_2$&$\phantom{-}0.285679$ & $ X_1Z_0Z_2$&$\phantom{-}0.348875$ \\
         $ X_2Z_0 $& $-0.221354$&$ Y_0Y_2$ &$ \phantom{-}0.142556$ &  $ X_1Y_0Y_2 $& $-0.175032$\\
         $ Z_0Z_2 $&$\phantom{-}0.279193$ &$ X_0X_1$ &$\phantom{-}0.123374$ & $ X_1X_2Z_0$&$-0.162602$ \\
         $ X_0Z_1 $&$\phantom{-}0.247090$ & $ X_1Z_0$&$\phantom{-}0.338724$ & $ X_0Z_1Z_2$&$-0.285679$\\
         $ Y_0Y_1  $&$\phantom{-}0.417991$ & $ Z_0Z_1$&$-0.583498$ & $ X_0X_2Z_1 $& $-1.479900$ \\
         $ X_0X_1X_2$&$-0.377952$ &$ X_0X_1Z_2$ & $-0.312326$ &$ X_0Y_1Y_2$ & $-0.930409$\\
        \hline
    \end{tabular}
\end{table*}

\begin{table*}[h]
    \centering
    \caption{The coefficients and tensor product operators of the Hamiltonian $H_{2^+}$ in GC scheme for $^{6}$Li.}
    \label{tab:H2+ligc}
    \renewcommand{\arraystretch}{1.2}
    \begin{tabular}{|c|c|c|c|c|c|}
        \hline
        \textbf{Operator} & \textbf{Coefficient} & \textbf{Operator} & \textbf{Coefficient}& \textbf{Operator} & \textbf{Coefficient} \\
        \hline
         $I$& $ -0.192300$& $X_1$ & $\phantom{-}0.871149$ & $ Y_0Y_1$&$\phantom{-}1.110249$ \\
         $ X_0$&$-0.871149$ & $ X_0X_1$&$-2.567049$&  $ Z_0Z_1$&$-1.139200$ \\
        \hline
    \end{tabular}
\end{table*}

\begin{table*}[h]
    \centering
    \caption{The coefficients and tensor product operators of the Hamiltonian $H_{3^+}$ in GC scheme for $^{6}$Li.}
    \label{tab:H3+ligc}
    \renewcommand{\arraystretch}{1.2}
    \begin{tabular}{|c|c|c|c|}
        \hline
        \textbf{Operator} & \textbf{Coefficient} & \textbf{Operator} & \textbf{Coefficient} \\
        \hline
         $I$& $ -2.5045$&$Z_0$ & $-2.5045 $\\
        \hline
    \end{tabular}
\end{table*}

\begin{table*}[h]
    \centering
    \caption{The coefficients and tensor product operators of the Hamiltonian $H_{0^+}$ in GC scheme for $^{38}$Ar.}
    \label{tab:H0+argc}
    \renewcommand{\arraystretch}{1.2}
    \begin{tabular}{|c|c|c|c|c|c|}
        \hline
        \textbf{Operator} & \textbf{Coefficient} & \textbf{Operator} & \textbf{Coefficient} & \textbf{Operator} & \textbf{Coefficient} \\
        \hline
        $I$  & $-93.974922$ & $X_2$  & $\phantom{-}0.179996$ & $X_0 X_1 X_2$&$\phantom{-}0.034904$ \\
        $ Z_2$&$-56.036235$ &$ X_1 $ &$\phantom{-}0.225232$ & $ X_0 X_1 Z_2$& $\phantom{-}0.034904$\\
        $ Z_1$&$19.260276$ & $ X_1 X_2$&$-0.143396$ & $ X_0 X_2 Z_1 $&$\phantom{-}0.034904$  \\
        $ Z_0 Z_2$&$18.679435$  & $ X_2 Z_1$& $-0.179996$ & $ X_0 Y_1 Y_2$&$-0.034904$ \\
        $ Y_1 Y_2$&$\phantom{-}0.143396$ & $ Z_1 Z_2$&$-18.678409$& $ X_0 Z_1 Z_2 $& $\phantom{-}0.034904$ \\
        $ X_0$&$-0.034904$ & $ Z_0$&$-19.259250$  & $ X_1 X_2 Z_0$&$-0.143396$ \\
        $ X_0 X_2$&$-0.034904$ &$ X_0 Z_2$ & $-0.034904$ & $ X_1 Y_0 Y_2 $&$\phantom{-}0.034904$ \\
        $ X_2 Z_0$& $\phantom{-}0.179996$&$ Y_0 Y_2$ &$-0.034904$ & $ X_1 Z_0 Z_2$&$\phantom{-}0.061559$ \\
        $ X_1 Z_2$&$\phantom{-}0.225232$&$ X_0 X_1$ &$ \phantom{-}0.034904$ & $ X_2 Y_0 Y_1 $&$\phantom{-}0.034904$\\
        $X_0 Z_1 $&$\phantom{-}0.034904$ & $ X_1 Z_0$&$\phantom{-}0.061559$  &$ X_2 Z_0 Z_1$ &$-0.179996$ \\
        $ Y_0 Y_1$&$\phantom{-}0.034904$ &$ Z_0 Z_1$ & $19.260280$ & $ Y_0 Y_1 Z_2 $& $\phantom{-}0.034904$ \\
        $ Y_1 Y_2 Z_0 $&$\phantom{-}0.143396$ & $ Z_0 Z_1 Z_2$&$-18.678405$  &$ Y_0 Y_2 Z_1$ &$\phantom{-}0.034904$ \\
        \hline
    \end{tabular}
\end{table*}

\begin{table*}[h]
    \centering
    \caption{The coefficients and tensor product operators of the Hamiltonian $H_{1^+}$ in GC scheme for $^{38}$Ar.}
    \label{tab:H1+argc}
    \renewcommand{\arraystretch}{1.2}
    \begin{tabular}{|c|c|c|c|c|c|}
        \hline
        \textbf{Operator} & \textbf{Coefficient} & \textbf{Operator} & \textbf{Coefficient} & \textbf{Operator} & \textbf{Coefficient} \\
        \hline
        $I$  & $ -112.474365$ & $X_1$  & $ \phantom{-}-0.120912$ &$ Y_0Y_1$&$\phantom{-}-0.141744$ \\
        $Z_1$  & $-37.676856$ & $ X_0$ & $ \phantom{-}\phantom{-}0.069808$ &$ Z_0Z_1$ &$ -36.956993$ \\
        $ Z_0$&$\phantom{-}37.840515$ &$ X_0X_1$ & $\phantom{-}\phantom{-} 0.141744$ & $ X_1Z_0$ &$\phantom{-}\phantom{-}0.120912$ \\
       $ X_0Z_1$ &$\phantom{-}\phantom{-}0.069808$ & &&&\\
        
        \hline
    \end{tabular}
\end{table*}

%

\begin{table*}[h]
    \centering
    \caption{The coefficients and tensor product operators of the Hamiltonian $H_{2^+}$ in GC scheme for $^{38}$Ar.}
    \label{tab:H2+argc}
    \renewcommand{\arraystretch}{1.2}
    \begin{tabular}{|c|c|c|c|c|c|}
        \hline
        \textbf{Operator} & \textbf{Coefficient} & \textbf{Operator} & \textbf{Coefficient} & \textbf{Operator} & \textbf{Coefficient} \\
        \hline
        $I$  & $ -150.396728$ & $X_0$  & $0.279234$ & $Z_0$  & $0.638015$ \\
        
        \hline
    \end{tabular}
\end{table*}

\begin{table*}[h]
    \centering
    \caption{The coefficients and tensor product operators of the Hamiltonian $H_{0^+}$ in JW scheme for $^{38}$Ar.}
    \label{tab:H0+arjw}
    \renewcommand{\arraystretch}{1.2}
    \begin{tabular}{|c|c|c|c|c|c|}
        \hline
        \textbf{Operator} & \textbf{Coefficient} & \textbf{Operator} & \textbf{Coefficient}&\textbf{Operator} & \textbf{Coefficient} \\
        \hline
        $I$  & $-187.949844$ & $X_0  X_1$  & $\phantom{-}\phantom{-}0.069808$&$ Z_1$& $\phantom{-}74.714645$ \\
        $ X_0 X_1 X_3 X_4$ &$\phantom{-}\phantom{-}0.112616$ &$ X_0 X_1 Y_3 Y_4$ &$\phantom{-}\phantom{-}0.112616$ &$ Z_1 Z_3$ & $-37.357837$\\
         $ X_0 Y_1 X_3 Y_4$&$\phantom{-}-0.030779$ &$ X_0 Y_1 Y_3 X_4$ &$\phantom{-}\phantom{-}0.030779$&$ Z_1 Z_4 $& $-37.356807$\\
         $ X_0 Z_1 X_2 X_3 Z_4 X_5$&$\phantom{-}\phantom{-}0.089998$ & $X_0 Z_1 X_2 Y_3 Z_4 Y_5$&$\phantom{-}\phantom{-}0.089998$ &$ Z_2$ & $\phantom{-}37.938686$\\
         $X_0 Z_1 X_2 X_4 X_5 $& $\phantom{-}-0.017452 $&  $ X_0 Z_1 X_2 Y_4 Y_5 $&$\phantom{-}-0.017452$ &$ Z_2 Z_5$& $-37.938686$  \\
         $ X_0 Z_1 Y_2 X_3 Z_4 Y_5$& $\phantom{-}-0.089998$&  $ X_0 Z_1 Y_2 Y_3 Z_4 X_5$ &$\phantom{-}0.089998$ & $ X_3 X_4$&$\phantom{-}-0.069808$ \\
         $ X_0 Z_1 Y_2 X_4 Y_5$&$\phantom{-}\phantom{-}0.017452$ & $X_0 Z_1 Y_2 Y_4 X_5 $&$\phantom{-}-0.017452$&$ Y_3 Y_4$& $ \phantom{-}-0.069808$ \\
         $ Y_0 X_1 X_3 Y_4$& $\phantom{-}\phantom{-}0.030779$&$ Y_0 X_1 Y_3 X_4$ & $\phantom{-}-0.030779$& $ Z_3$&$\phantom{-}75.296516$ \\
         $ Y_0 Y_1$&$\phantom{-}\phantom{-}0.069808$& $ Y_0 Y_1 X_3 X_4$&$\phantom{-}\phantom{-}0.112616$&$ Z_4$&$\phantom{-}74.714641$  \\
         $ Y_0 Y_1 Y_3 Y_4$& $\phantom{-}\phantom{-}0.112616$& $ Y_0 Z_1 X_2 X_3 Z_4 Y_5$&$\phantom{-}\phantom{-}0.089998$ &$ Z_5$ &$\phantom{-}37.938686$  \\
         $ Y_0 Z_1 X_2 Y_4 X_5$& $\phantom{-}\phantom{-}0.017452$& $ Y_0 Z_1 Y_2 X_3 Z_4 X_5$&$\phantom{-}\phantom{-}0.089998$ & $ Z_0$&$\phantom{-}75.296512$ \\
         $ Y_0 Z_1 Y_2 Y_3 Z_4 Y_5$& $\phantom{-}\phantom{-}0.089998$& $ Y_0 Z_1 Y_2 X_4 X_5$ & $\phantom{-}-0.017452$&$ Z_0 Z_3$&$-37.938678$ \\
         $ X_1 X_2 X_3 Z_4 X_5$&$\phantom{-}\phantom{-}0.017452$ & $ X_1 X_2 Y_3 Z_4 Y_5$&$\phantom{-}\phantom{-}0.017452$ &$ Z_0 Z_4$ & $-37.357833$\\
         $ X_1 X_2 X_4 X_5 $&$\phantom{-}-0.071698$ & $ X_1 X_2 Y_4 Y_5$&$\phantom{-}-0.071698$ &$ Y_1 Y_2 X_3 Z_4 X_5$ &$\phantom{-}\phantom{-}0.017452$ \\
          $ X_1 Y_2 X_3 Z_4 Y_5$& $\phantom{-}-0.017452$&$ X_1 Y_2 Y_3 Z_4 X_5$ &$\phantom{-}\phantom{-}0.017452$ &$ Y_1 Y_2 Y_3 Z_4 Y_5$ & $\phantom{-}\phantom{-}0.017452$\\
         $ X_1 Y_2 X_4 Y_5$& $\phantom{-}\phantom{-}0.071698$& $ X_1 Y_2 Y_4 X_5$& $\phantom{-}-0.071698$&$ Y_1 Y_2 X_4 X_5$& $\phantom{-}-0.071698$\\
          $ Y_1 X_2 X_3 Z_4 Y_5$&$\phantom{-}\phantom{-}0.017452$ &$ Y_1 X_2 Y_3 Z_4 X_5$ & $\phantom{-}-0.017452$&$ Y_1 Y_2 Y_4 Y_5$ & $\phantom{-}-0.071698$\\
          $ Y_0 Z_1 X_2 Y_3 Z_4 X_5$& $\phantom{-}-0.089998$& $ Y_0 Z_1 X_2 X_4 Y_5$ & $\phantom{-}-0.017452$&$ Y_0 Z_1 Y_2 Y_4 Y_5$&$\phantom{-}-0.017452$\\
          $ Y_1 X_2 X_4 Y_5$&$\phantom{-}-0.071698$ &$ Y_1 X_2 Y_4 X_5$ & $\phantom{-}\phantom{-}0.071698$&&\\        
        \hline
    \end{tabular}
\end{table*}

\begin{table*}[h]
    \centering
    \caption{The coefficients and tensor product operators of the Hamiltonian $H_{1^+}$ in JW scheme for $^{38}$Ar.}
    \label{tab:H1+arjw}
    \renewcommand{\arraystretch}{1.2}
    \begin{tabular}{|c|c|c|c|c|c|}
        \hline
        \textbf{Operator} & \textbf{Coefficient} & \textbf{Operator} & \textbf{Coefficient} & \textbf{Operator} & \textbf{Coefficient} \\
        \hline
        $I$  & $ -112.474365$ & $X_0  X_1$  & $\phantom{-}\phantom{-}0.069808$& $Y_0  Y_1$ & $\phantom{-}0.069808$  \\
        $Y_0  Z_1  Z_2  Y_3  Y_4  Y_5$ & $-0.030228$ & $ X_1  Z_2  X_3  X_4  X_5$ & $\phantom{-}0.035436$ & $ Z_1  Z_5 $& $-37.316925$\\
        $X_0  Z_1  Z_2  X_3  X_4  X_5$ & $-0.030228$ & $X_0  Z_1  Z_2  X_3  Y_4  Y_5$ & $-0.030228$ & $Z_3 $&$\phantom{-}37.398754$  \\
        $X_0  Z_1  Z_2  Y_3  X_4  Y_5$ & $-0.030228$ & $X_0  Z_1  Z_2  Y_3  Y_4  X_5$ & $\phantom{-}0.030228$& $Z_3  Z_4$& $-37.398754$\\
        $ Y_1  Z_2  Y_3  Y_4  Y_5$& $\phantom{-}0.035436$ & $Y_0  Z_1  Z_2  X_3 X_4  Y_5$ & $\phantom{-}0.030228$ & $Z_4 $&$\phantom{-}37.398754$ \\
        $Y_0  Z_1  Z_2  X_3  Y_4  X_5$ & $-0.030228$ & $Y_0  Z_1  Z_2  Y_3  X_4  X_5$ & $-0.030228$& $ Z_5$&$\phantom{-}75.075611$ \\
        $ X_1  Z_2  X_3  Y_4  Y_5 $ & $\phantom{-}0.035436$&$ X_1  Z_2  Y_3  X_4  Y_5$ & $\phantom{-}0.035436$&$ Z_0  Z_5 $ & $-37.758686$ \\
        $ X_1  Z_2  Y_3  Y_4  X_5$&$-0.035436$ &$ Y_1  Z_2  X_3  X_4  Y_5$ & $-0.035436$& $Z_0 $ & $\phantom{-}37.758686$ \\
        $ Y_1  Z_2  X_3  Y_4  X_5$&$\phantom{-}0.035436$ &$ Y_1  Z_2  Y_3  X_4  X_5$ & $\phantom{-}0.035436$ & $ Z_1 $&$\phantom{-}37.316925$ \\

        \hline
    \end{tabular}
\end{table*}

\begin{table*}[h]
    \centering
    \caption{The coefficients and tensor product operators of the Hamiltonian $H_{2^+}$ in JW scheme for $^{38}$Ar.}
    \label{tab:H2+arjw}
    \renewcommand{\arraystretch}{1.2}
    \begin{tabular}{|c|c|c|c|c|c|}
        \hline
        \textbf{Operator} & \textbf{Coefficient} & \textbf{Operator} & \textbf{Coefficient}  & \textbf{Operator} & \textbf{Coefficient} \\
        \hline
        $I$  & $-75.198364$ & $Z_3 Z_5$  & $-37.439678$ & $Z_3$  & $\phantom{-}37.439678$\\
        $X_3 X_4$  & $\phantom{-}\phantom{-}0.139617$ & $Z_4$  & $37.758686$ & $Z_5$  & $75.198364$ \\
        $Y_3 Y_4$  & $\phantom{-}\phantom{-}0.139617$ & $Z_4 Z_5$  & $-37.758686$ & &\\
         
        \hline
    \end{tabular}
\end{table*}

\section{Ansätze}\label{app:result_ansatz}
This section provides the explicit ansätze corresponding to the two algorithms used in our quantum simulations.

\subsection{Qubit-ADAPT-VQE}\label{app:ansatz_ADAPT}
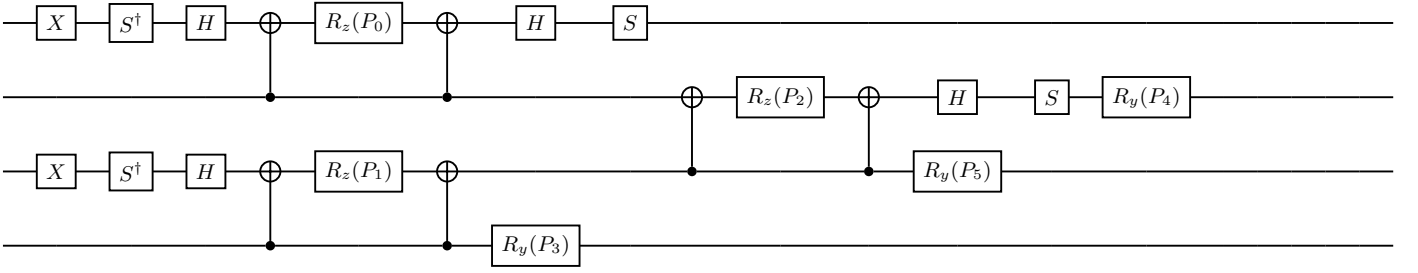
\begin{figure*}
\begin{tikzpicture}
\node[scale=0.9] {
\begin{quantikz}
\qw & \gate{X} & \gate{S^\dagger} & \gate{H}&\targ{} &\gate{R_z(P_0)}&\targ{}& \gate{H} & \gate{S} & \qw&\qw & \qw &\qw&\qw & \qw & \qw & \qw & \qw & \qw & \qw & \qw  \\
\qw & \qw & \qw & \qw & \ctrl{-1} &\qw & \ctrl{-1}& \qw & \qw&\targ{} &\gate{R_z(P_2)}&\targ{}& \gate{H} & \gate{S}  & \gate{R_y(P_4)}& \qw & \qw & \qw & \qw & \qw & \qw   \\
\qw & \gate{X} & \gate{S^\dagger} & \gate{H}&\targ{} &\gate{R_z(P_1)}&\targ{}& \qw & \qw & \ctrl{-1} &\qw & \ctrl{-1}& \gate{R_y(P_5)}&\qw & \qw & \qw & \qw & \qw & \qw & \qw & \qw   \\
\qw & \qw & \qw & \qw & \ctrl{-1} &\qw & \ctrl{-1}& \gate{R_y(P_3)} & \qw & \qw & \qw  & \qw & \qw &\qw & \qw & \qw & \qw & \qw & \qw & \qw & \qw  \\
\end{quantikz}
};
\end{tikzpicture}
\caption{Quantum circuit for $0^+$ state of $^{6}$Li using qubit ADAPT-VQE in GC scheme. }\label{fig:adapt0+ligc}
\end{figure*}

\begin{figure*}
\begin{tikzpicture}
\node[scale=1.0] {
\begin{quantikz}
\qw & \gate{X} & \gate{S^\dagger} & \gate{H} & \qw&\qw & \qw &\qw&\qw & \targ{} &\gate{R_z(P_1)}&\targ{}& \gate{H} & \gate{S} & \gate{R_y(P_3)} & \qw    \\
\qw & \gate{X} & \gate{S^\dagger} & \gate{H}&\targ{} &\gate{R_z(P_0)}&\targ{}& \gate{H} & \gate{S}& \ctrl{-1} &\qw & \ctrl{-1}& \qw & \qw & \qw & \qw    \\
\qw & \qw & \qw & \qw & \ctrl{-1} &\qw & \ctrl{-1}& \gate{R_y(P_2)} & \qw & \qw & \qw  & \qw & \qw &\qw & \qw & \qw \\
\end{quantikz}
};
\end{tikzpicture}
\caption{Quantum circuit for $1^+$ state of $^{6}$Li using qubit ADAPT-VQE in GC scheme. }\label{fig:adapt1+ligc}
\end{figure*}
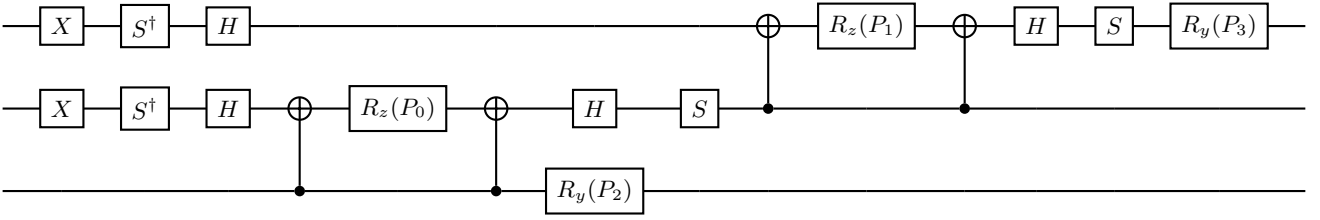

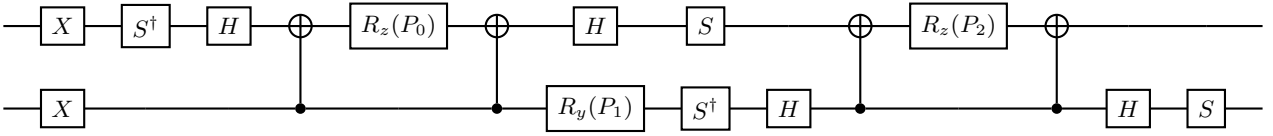
\begin{figure*}
\begin{tikzpicture}
\node[scale=1.0] {
\begin{quantikz}
\qw & \gate{X} & \gate{S^\dagger} & \gate{H} & \targ{} &\gate{R_z(P_0)}&\targ{}& \gate{H} & \gate{S} & \qw & \targ{} &\gate{R_z(P_2)}&\targ{}& \qw & \qw & \qw   \\
\qw & \gate{X} & \qw & \qw & \ctrl{-1} &\qw & \ctrl{-1}&\gate{R_y(P_1)} &\gate{S^\dagger} & \gate{H} &\ctrl{-1} &\qw & \ctrl{-1} &\gate{H} & \gate{S} & \qw    \\
\end{quantikz}
};
\end{tikzpicture}
\caption{Quantum circuit for $2^+$ state of $^{6}$Li using qubit ADAPT-VQE in GC scheme. }\label{fig:adapt2+ligc}
\end{figure*}

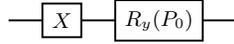
\begin{figure*}
\begin{tikzpicture}
\node[scale=0.9] {
\begin{quantikz}
\qw & \gate{X}  & \gate{R_y(P_0)} & \qw \\
\end{quantikz}
};
\end{tikzpicture}
\caption{Quantum circuit for $3^+$ state of $^{6}$Li and $2^+$ state of $^{38}$Ar using qubit ADAPT-VQE in GC scheme. }\label{fig:adapt3+ligc}
\end{figure*}

\begin{figure*}
\begin{tikzpicture}
\node[scale=0.8] {
\begin{quantikz}
\qw& \gate{X}  & \gate{S^\dagger} & \gate{H}&\qw& \qw &\qw & \qw & \qw  & \qw &\targ{} &\gate{R_z(P_2)}&\targ{}& \gate{H} &\gate{S} & \gate{R_y(P_4)} & \qw &\qw \\
\qw & \gate{X} & \gate{S^\dagger} & \gate{H}&\targ{} &\gate{R_z(P_0)}&\targ{}& \gate{H} & \gate{S} & \gate{R_y(P_1)} &\ctrl{-1} & \qw &\ctrl{-1}&\qw & \qw & \qw &\qw & \qw  \\
\qw & \qw & \qw & \qw & \ctrl{-1} &\qw & \ctrl{-1}& \qw & \qw & \qw & \qw  & \qw & \qw &\qw & \qw & \qw &\qw & \qw \\
\end{quantikz}
};
\end{tikzpicture}
\caption{Quantum circuit for $0^+$ state of $^{38}$Ar using qubit ADAPT-VQE in GC scheme. }\label{fig:adapt0+argc}
\end{figure*}
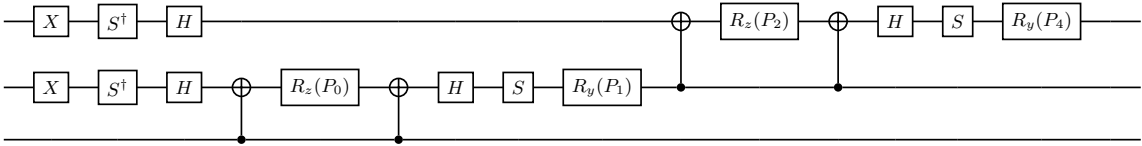

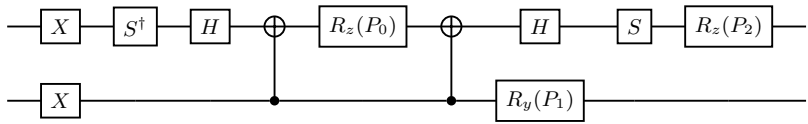
\begin{figure*}
\begin{tikzpicture}
\node[scale=0.9] {
\begin{quantikz}
\qw &\gate{X} & \gate{S^\dagger} & \gate{H}&\targ{} &\gate{R_z(P_0)}&\targ{}& \gate{H} & \gate{S} & \gate{R_z(P_2)} & \qw\\
\qw & \gate{X}  & \qw & \qw & \ctrl{-1} &\qw & \ctrl{-1}& \gate{R_y(P_1)} & \qw & \qw & \qw \\
\end{quantikz}
};
\end{tikzpicture}
\caption{Quantum circuit for $1^+$ state of $^{38}$Ar using qubit ADAPT-VQE in GC scheme. }\label{fig:adapt1+argc}
\end{figure*}

\begin{figure*}
\begin{tikzpicture}
\node[scale=0.5] {
\begin{quantikz}
\qw  &  \gate{S^\dagger} & \gate{H} &\targ{} &\gate{R_z(P_0)}&\targ{} & \gate{H} & \gate{S}& \qw & \qw & \qw & \qw & \qw &\qw & \qw & \qw & \qw & \qw & \qw  & \qw & \qw & \qw & \qw & \qw & \qw  &\qw& \qw & \qw & \qw & \qw  \\
\qw & \qw& \qw & \ctrl{-1} &\qw & \ctrl{-1}&\gate{S^\dagger} & \gate{H}& \targ{} &\gate{R_z(P_1)}&\targ{} & \gate{H} & \gate{S}&\qw & \qw & \qw & \qw & \qw & \qw  & \qw & \qw & \qw&\qw &\qw &\qw&\qw&\qw & \qw& \qw &  \qw    \\
\qw &\qw &\qw & \qw & \qw &\qw & \qw & \qw & \ctrl{-1} &\qw & \ctrl{-1}&\gate{S^\dagger} & \gate{H}&\targ{} &\gate{R_z(P_2)}&\targ{} & \gate{H} & \gate{S}&\qw& \qw &\qw& \qw &  \qw  &\qw & \qw &\qw& \qw & \qw & \qw & \qw   \\
\qw & \gate{x} & \qw&\qw& \qw &\qw & \qw & \qw & \qw & \qw &\qw &\qw&\qw& \ctrl{-1} &\qw & \ctrl{-1}&\gate{S^\dagger} & \gate{H}&\targ{} &\gate{R_z(P3)}&\targ{} & \gate{H} & \gate{S}& \qw  & \qw & \qw  & \qw& \qw   & \qw & \qw   \\
\qw & \qw & \qw& \qw&\qw &\qw&\qw& \qw &\qw & \qw &\qw & \qw &\qw&\qw & \qw & \qw & \qw & \qw &  \ctrl{-1} &\qw & \ctrl{-1}&\gate{S^\dagger} & \gate{H}& \targ{} &\gate{R_z(P4)}&\targ{} & \gate{H} & \gate{S}& \gate{R_y(P5)} & \qw   \\
\qw &\gate{x}& \qw & \qw & \qw &\qw & \qw& \qw & \qw & \qw & \qw  & \qw & \qw &\qw & \qw & \qw & \qw & \qw & \qw & \qw & \qw & \qw & \qw & \ctrl{-1} &\qw & \ctrl{-1}&\gate{S^\dagger} & \gate{H}& \gate{R_y(P6)}& \qw  \\
\end{quantikz}
};
\end{tikzpicture}
\caption{Quantum circuit for $2^+$ state of $^{38}$Ar using qubit ADAPT-VQE in JW scheme. }\label{fig:adapt2+arjw}
\end{figure*}
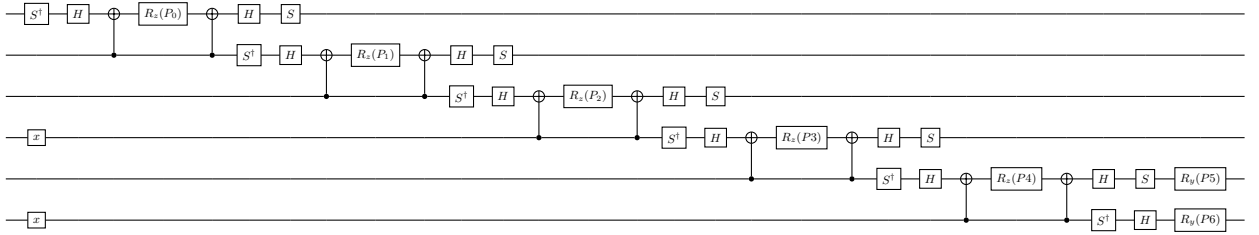

The final ansätze obtained using the qubit-ADAPT-VQE (Section~\ref{sec:qubit_adapt}) for various spin-parity states of $^{38}$Ar and $^{6}$Li are given Figs.~\ref{fig:adapt0+ligc}, ~\ref{fig:adapt1+ligc}, ~\ref{fig:adapt2+ligc}, ~\ref{fig:adapt3+ligc}, ~\ref{fig:adapt0+argc}, ~\ref{fig:adapt1+argc}, ~\ref{fig:adapt0+arjw}, ~\ref{fig:adapt1+arjw}, and~\ref{fig:adapt2+arjw}. These ansätze are built iteratively using the pre-defined operator pool. 

\begin{figure*}
\begin{tikzpicture}
\node[scale=0.4] {
\begin{quantikz}
\qw & \gate{R_y(P_8)} & \gate{S^\dagger} & \gate{H}& \qw &\targ{} &\gate{R_z(P_5)}&\targ{}& \gate{H} & \gate{S} & \qw & \qw & \qw &\qw & \qw & \qw & \qw & \qw & \qw  & \qw & \qw & \qw & \qw & \qw & \qw  & \qw & \qw & \qw & \qw   &\targ{} &\gate{R_z(P_10)}&\targ{}& \gate{H} & \gate{S} & \qw \\
\qw & \gate{X}& \gate{S^\dagger} & \gate{H}& \qw  &\ctrl{-1} & \qw &\ctrl{-1} & \qw & \qw & \qw & \qw & \qw &\qw & \qw & \qw & \qw & \qw & \qw  & \qw & \qw & \qw &\targ{} &\gate{R_z(P_7)}&\targ{}& \gate{H} & \gate{S}& \qw &  \gate{R_y(P_8)}  &\ctrl{-1} & \qw &\ctrl{-1}& \qw & \qw & \qw  \\
\qw & \gate{S^\dagger} & \gate{H} & \qw & \qw &\qw & \qw & \qw & \qw & \qw & \qw & \qw & \qw &\qw & \qw &\targ{} &\gate{R_z(P_4)}&\targ{}& \gate{H} & \gate{S}& \qw &  \gate{R_y(P_6)}  &\ctrl{-1} & \qw &\ctrl{-1}& \qw & \qw & \qw & \qw & \qw  & \qw & \qw & \qw & \qw & \qw  \\
\qw & \gate{S^\dagger} & \gate{H}&\qw& \qw &\qw & \qw & \qw & \qw & \qw &\targ{} &\gate{R_z(P_3)}&\targ{}& \gate{H} & \gate{S}&\ctrl{-1} & \qw &\ctrl{-1}& \qw &\gate{R_y(P_11)} &  \qw & \qw & \qw  & \qw & \qw  & \qw& \qw   & \qw & \qw & \qw  & \qw & \qw & \qw & \qw & \qw  \\
\qw & \gate{X} & \gate{S^\dagger} & \gate{H}&\targ{} &\gate{R_z(P_0)}&\targ{}& \gate{H} & \gate{S} & \gate{R_y(P_1)} &\ctrl{-1} & \qw &\ctrl{-1}&\qw & \qw & \qw & \qw & \qw & \qw & \qw & \qw & \qw & \qw & \qw & \qw  & \qw & \qw & \qw & \qw &\targ{} &\gate{R_z(P_12)}&\targ{}& \gate{H} & \gate{S}  & \qw  \\
\qw & \qw & \qw & \qw & \ctrl{-1} &\qw & \ctrl{-1}& \gate{R_y(P_2)} & \qw & \qw & \qw  & \qw & \qw &\qw & \qw & \qw & \qw & \qw & \qw & \qw & \qw & \qw & \qw & \qw & \qw  & \qw & \qw & \qw & \qw&\ctrl{-1} & \qw &\ctrl{-1} & \qw & \qw & \qw  \\
\end{quantikz}
};
\end{tikzpicture}
\caption{Quantum circuit for $0^+$ state of $^{38}$Ar using qubit ADAPT-VQE in JW scheme. }\label{fig:adapt0+arjw}
\end{figure*}
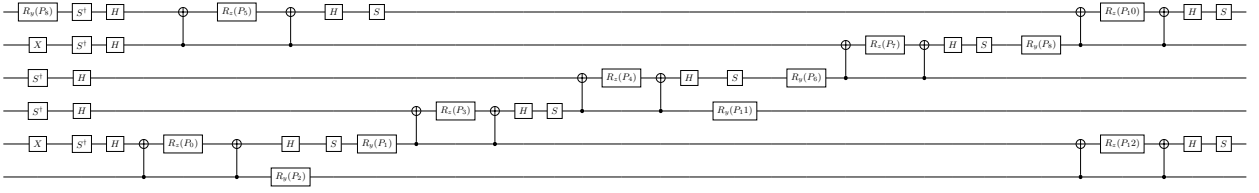

\begin{figure*}
\begin{tikzpicture}
\node[scale=0.4] {
\begin{quantikz}
\qw  & \gate{S^\dagger} & \gate{H}& \qw &\qw&\qw  & \qw & \qw & \qw & \qw & \qw & \qw & \qw &\qw & \qw & \qw & \qw & \qw & \qw  & \qw & \qw & \qw & \qw & \qw & \qw  &\qw& \qw & \qw & \qw & \qw   &\targ{} &\gate{R_z(P_8)}&\targ{}& \gate{H} & \gate{S} & \qw \\
\qw & \gate{X}& \gate{S^\dagger} & \gate{H}& \qw &\qw & \qw & \qw & \qw & \qw & \qw & \qw & \qw &\qw & \qw & \qw & \qw & \qw & \qw  & \qw & \qw & \qw&\qw &\targ{} &\gate{R_z(P_6)}&\targ{}& \gate{H} & \gate{S}& \qw &  \gate{R_y(P_7)}  &\ctrl{-1} & \qw &\ctrl{-1}& \qw & \qw & \qw  \\
\qw & \gate{S^\dagger} & \gate{H} & \qw & \qw &\qw & \qw & \qw & \qw & \qw & \qw & \qw & \qw &\qw & \qw&\qw &\targ{} &\gate{R_z(P_4)}&\targ{}& \gate{H} & \gate{S}& \qw &  \gate{R_y(P_5)}  &\ctrl{-1} & \qw &\ctrl{-1}& \qw & \qw & \qw & \qw & \qw  & \qw & \qw & \qw & \qw & \qw  \\
\qw & \gate{S^\dagger} & \gate{H}&\qw& \qw &\qw & \qw & \qw & \qw & \qw &\targ{} &\gate{R_z(P_2)}&\targ{}& \gate{H} & \gate{S}& \gate{R_y(P_3)}&\ctrl{-1} & \qw &\ctrl{-1}& \qw &\qw &  \qw & \qw & \qw  & \qw & \qw  & \qw& \qw   & \qw & \qw & \qw  & \qw & \qw & \qw & \qw & \qw  \\
\qw & \gate{S^\dagger} & \gate{H}& \qw&\targ{} &\gate{R_z(P_0)}&\targ{}& \gate{H} & \gate{S} & \gate{R_y(P_1)} &\ctrl{-1} & \qw &\ctrl{-1}&\qw & \qw & \qw & \qw & \qw & \qw & \qw & \qw & \qw & \qw & \qw & \qw  & \qw & \qw & \qw & \qw & \qw & \qw  & \qw & \qw &\qw & \qw &\qw  \\
\qw & \gate{X}& \qw & \qw & \ctrl{-1} &\qw & \ctrl{-1}& \qw & \qw & \qw & \qw  & \qw & \qw &\qw & \qw & \qw & \qw & \qw & \qw & \qw & \qw & \qw & \qw & \qw & \qw  & \qw & \qw & \qw & \qw & \qw &  \qw  & \qw & \qw & \qw & \qw &\qw  \\
\end{quantikz}
};
\end{tikzpicture}
\caption{Quantum circuit for $1^+$ state of $^{38}$Ar using qubit ADAPT-VQE in JW scheme. }\label{fig:adapt1+arjw}
\end{figure*}
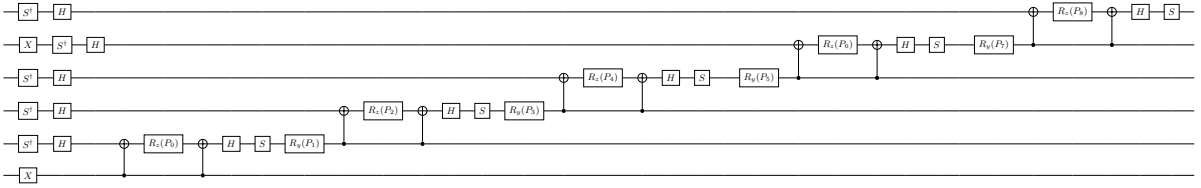

\subsection{VQE}\label{app:ansatz_VQE}
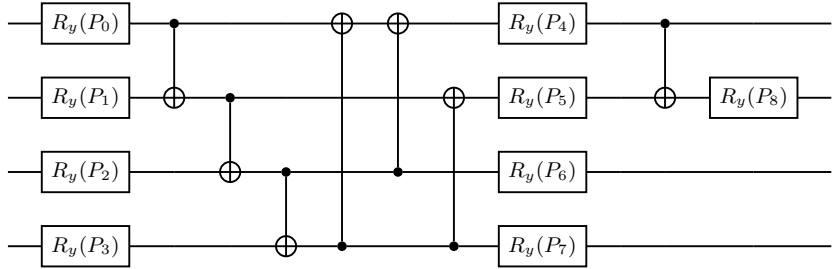
\begin{figure*}
\begin{tikzpicture}
\node[scale=0.9] {
\begin{quantikz}
\qw & \gate{R_y(P_0)} & \ctrl{1} & \qw      & \qw      & \targ{} & \targ{}&\qw & \gate{R_y(P_4)} & \qw      & \ctrl{1} & \qw & \qw \\
\qw & \gate{R_y(P_1)} & \targ{}  & \ctrl{1} & \qw      & \qw      & \qw &\targ{}& \gate{R_y(P_5)} & \qw      & \targ{}  & \gate{R_y(P_8)} & \qw \\
\qw & \gate{R_y(P_2)} & \qw      & \targ{}  & \ctrl{1} & \qw  & \ctrl{-2} &\qw     & \gate{R_y(P_6)} & \qw      & \qw      & \qw & \qw \\
\qw & \gate{R_y(P_3)} & \qw      & \qw      & \targ{}  & \ctrl{-3}      & \qw&\ctrl{-2}  & \gate{R_y(P_7)} & \qw      & \qw      & \qw & \qw\\
\end{quantikz}
};
\end{tikzpicture}      
\caption{Quantum circuit for $0^+$ state of $^{6}$Li for VQE in GC scheme. }\label{fig:adapt0+ligcvqe}
\end{figure*}

\begin{figure*}
\begin{tikzpicture}
\node[scale=0.9] {
\begin{quantikz}
\qw  & \gate{R_y(P_0)} & \qw \\
\end{quantikz}
};
\end{tikzpicture}
\caption{Quantum circuit for $3^+$ state of  $^{6}$Li, $2^+$ state of $^{38}$Ar for VQE in GC scheme. }\label{fig:vqe3+ligc}
\end{figure*}
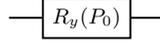

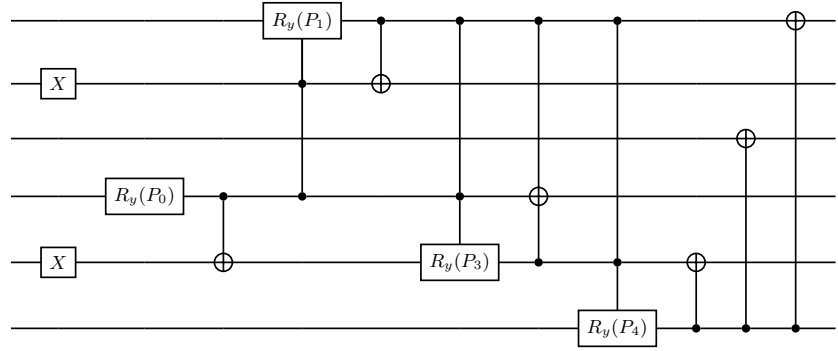
\begin{figure*}
\begin{tikzpicture}
\node[scale=0.8] {
\begin{quantikz}
\qw & \qw  & \qw& \qw   & \gate{R_y(P_1)} &  \ctrl{1} & \ctrl{3}  &  \ctrl{3} &  \ctrl{4}& \qw & \qw & \targ{} & \qw  \\
\qw & \gate{X}  & \qw& \qw   & \ctrl{-1} & \targ{} & \qw  & \qw & \qw & \qw & \qw & \qw  & \qw  \\
\qw & \qw  & \qw& \qw   & \qw & \qw & \qw  & \qw & \qw & \qw & \targ{} & \qw & \qw  \\
\qw & \qw  &\gate{R_y(P_0)}& \ctrl{1}   &  \ctrl{-3} & \qw & \ctrl{1}  & \targ{} & \qw & \qw & \qw & \qw & \qw  \\
\qw & \gate{X}  & \qw& \targ{}   & \qw & \qw & \gate{R_y(P_3)}  & \ctrl{-1} & \ctrl{1} & \targ{} & \qw & \qw & \qw  \\
\qw & \qw  & \qw& \qw   & \qw & \qw & \qw  & \qw &  \gate{R_y(P_4)} & \ctrl{-1} & \ctrl{-3} & \ctrl{-5} & \qw  \\
\end{quantikz}
};
\end{tikzpicture}
\caption{Quantum circuit for $0^+$ state of $^{38}$Ar for VQE in JW scheme. }\label{fig:vqe0+arjw}
\end{figure*}

\begin{figure*}
\begin{tikzpicture}
\node[scale=0.8] {
\begin{quantikz}
\qw  & \qw& \gate{R_y(P_0)}   & \ctrl{1} & \ctrl{3} & \targ{}  & \qw & \qw & \qw \\
\qw & \gate{X}& \qw   & \targ{} & \qw & \qw  & \qw & \qw & \qw   \\
\qw  & \qw& \qw   & \qw & \qw & \qw  & \qw & \qw & \qw  \\
\qw  & \qw& \qw   & \qw & \gate{R_y(P_1)} & \ctrl{-3}  & \ctrl{1} & \ctrl{2}& \qw   \\
\qw  & \qw& \qw   & \qw & \qw & \qw  & \targ{} & \qw & \qw   \\
\qw & \gate{X}& \qw   & \qw & \qw & \qw  & \qw & \targ{} & \qw   \\
\end{quantikz}
};
\end{tikzpicture}
\caption{Quantum circuit for $1^+$ state of $^{38}$Ar for VQE in JW scheme. }\label{fig:vqe1+arjw}
\end{figure*}
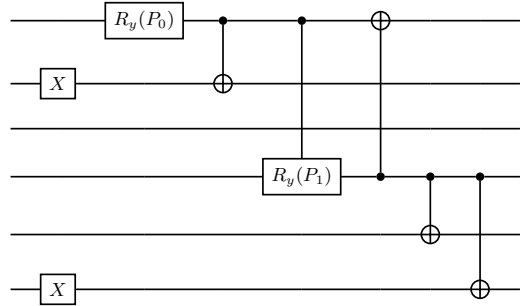

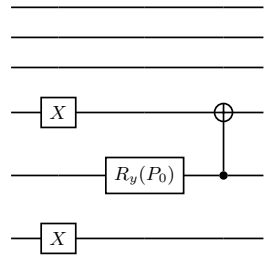
\begin{figure*}
\begin{tikzpicture}
\node[scale=0.8] {
\begin{quantikz}
\qw & \qw      & \qw      & \qw      & \qw      \\
\qw & \qw      & \qw      & \qw      & \qw      \\
\qw & \qw      & \qw      & \qw      & \qw      \\
\qw & \gate{X} & \qw      & \targ{} & \qw      \\
\qw & \qw      & \gate{R_y(P_0)} & \ctrl{-1}  & \qw      \\
\qw & \gate{X} & \qw      & \qw      & \qw\\
\end{quantikz}
};
\end{tikzpicture}
\caption{Quantum circuit for $2^+$ state of $^{38}$Ar for VQE in JW scheme. }\label{fig:vqe2+arjw}
\end{figure*}

Here we have given the ansätze used for VQE. For JW scheme the given ansätze are prepared using the particle conserving ansatz technique given in appendix~\ref{app:ansatz}. While for the GC encoding hardware efficient variational ansätze are used. The corresponding figures are shown in Figs.~\ref{fig:adapt0+ligcvqe}, ~\ref{fig:vqe1+ligc}, ~\ref{fig:vqe2+ligc}, ~\ref{fig:vqe3+ligc},~\ref{fig:vqe0+arjw},~\ref{fig:vqe1+arjw}, and~\ref{fig:vqe2+arjw}. 

\begin{figure*}
\begin{tikzpicture}
\node[scale=1.0] {
\begin{quantikz}
\qw&\gate{R_y(P_0)} & \ctrl{1} & \qw   & \targ{}   & \gate{R_y(P_3)} & \ctrl{1}     & \qw & \qw \\
\qw&\gate{R_y(P_1)} & \targ{}  & \ctrl{1}&\qw & \gate{R_y(P_4)} & \targ{} & \gate{R_y(P_6)}  & \qw \\
\qw&\gate{R_y(P_2)} & \qw      & \targ{} &\ctrl{-2} & \gate{R_y(P_5)} & \qw      & \qw      & \qw\\
\end{quantikz}
};
\end{tikzpicture}
\caption{Quantum circuit for $1^+$ state of $^{6}$Li and $0^+$ state of $^{38}$Ar for VQE in GC scheme. }\label{fig:vqe1+ligc}
\end{figure*}
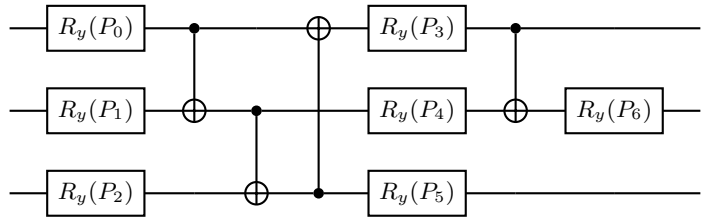

\begin{figure*}
\begin{tikzpicture}
\node[scale=1.0] {
\begin{quantikz}
\qw & \gate{R_y(P_0)} & \ctrl{1} & \qw      & \qw \\
\qw & \gate{R_y(P_1)} & \targ{}  & \gate{R_y(P_2)} & \qw
\end{quantikz}
};
\end{tikzpicture}
\caption{Quantum circuit for $2^+$ state of $^{6}$Li and $1^+$ state of $^{38}$Ar for VQE in GC scheme. }\label{fig:vqe2+ligc}
\end{figure*}
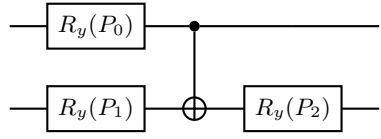

\end{appendix}
\bibliography{refrences}
\end{document}